\documentclass[11pt]{article}

\usepackage[final]{ACL2023}

\usepackage{times}
\usepackage{latexsym}

\usepackage[T1]{fontenc}
\usepackage[utf8]{inputenc}
\usepackage{soul}

\usepackage{microtype}

\usepackage{inconsolata}
\usepackage[table,xcdraw]{xcolor}

\usepackage{algpseudocode}
\usepackage{algorithm}

\usepackage{booktabs} 

\usepackage{graphicx}
\usepackage{amsmath}
\usepackage{mathtools, amssymb, lipsum, multicol}
\usepackage{soul}
\usepackage{todonotes}
\usepackage{array}
\usepackage{hyperref}
\usepackage{multirow}
\usepackage{makecell}
\usepackage{scalerel}
\usepackage{enumitem}

\usepackage[most]{tcolorbox} % Loads tcolorbox with most features
\usepackage{xcolor}          % For coloring the box (optional but recommended)

\tcbset{colback=gray!10, colframe=gray!50, coltitle=white, 
fonttitle=\bfseries, sharp corners, fontupper=\small}

\usepackage{float} 
\usepackage{fontawesome5}
\newlength{\myMheight}
\makeatletter
\let\old@footnotemark\@footnotemark
\renewcommand{\@footnotemark}{\hbox{\githubmarker}\old@footnotemark}
\makeatother

\title{GRITHopper: Decomposition-Free Multi-Hop Dense Retrieval}
\author{
\hspace{-.7em}Justus-Jonas Erker\hspace{.05em}\scalerel*{\includegraphics{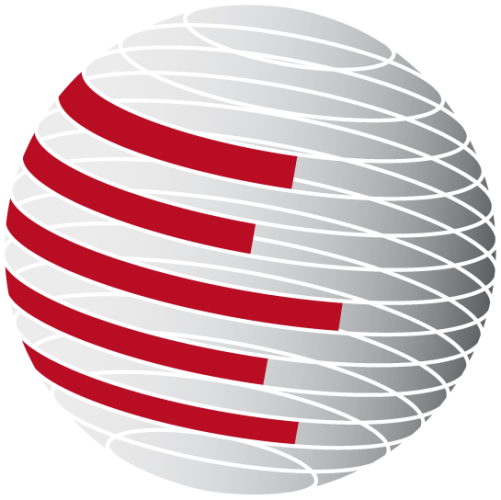}}{\textbf{O}}\hspace{.1em} 
Nils Reimers\hspace{.05em}\scalerel*{\includegraphics{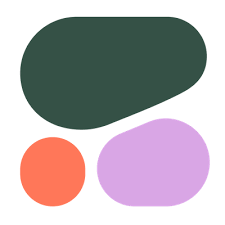}}{\textbf{O}}\hspace{.1em}
Iryna Gurevych\hspace{.05em}\scalerel*{\includegraphics{images/2010-06-04_logo-ukp_gross_weisser-hintergrund_0x500.png}}{\textbf{O}} \\
\hspace{-.4em}\scalerel*{\includegraphics{images/2010-06-04_logo-ukp_gross_weisser-hintergrund_0x500.png}}{\textbf{O}} Ubiquitous Knowledge Processing Lab (UKP Lab) \\
     Department of Computer Science and Hessian Center for AI (hessian.AI) \\
     Technical University of Darmstadt \\
    \url{www.ukp.tu-darmstadt.de} \\ \scalerel* {\includegraphics{images/cohere.png}}{\textbf{O}} cohere \hfill
}

\newcommand{\githublogo}{\protect\raisebox{-1pt}{\includegraphics[height=0.35cm]{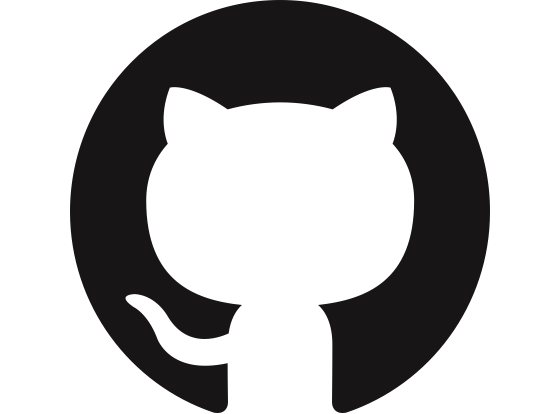}}}
\newcommand{\huggingfacelogo}{\protect\raisebox{-1pt}{\includegraphics[height=0.32cm]{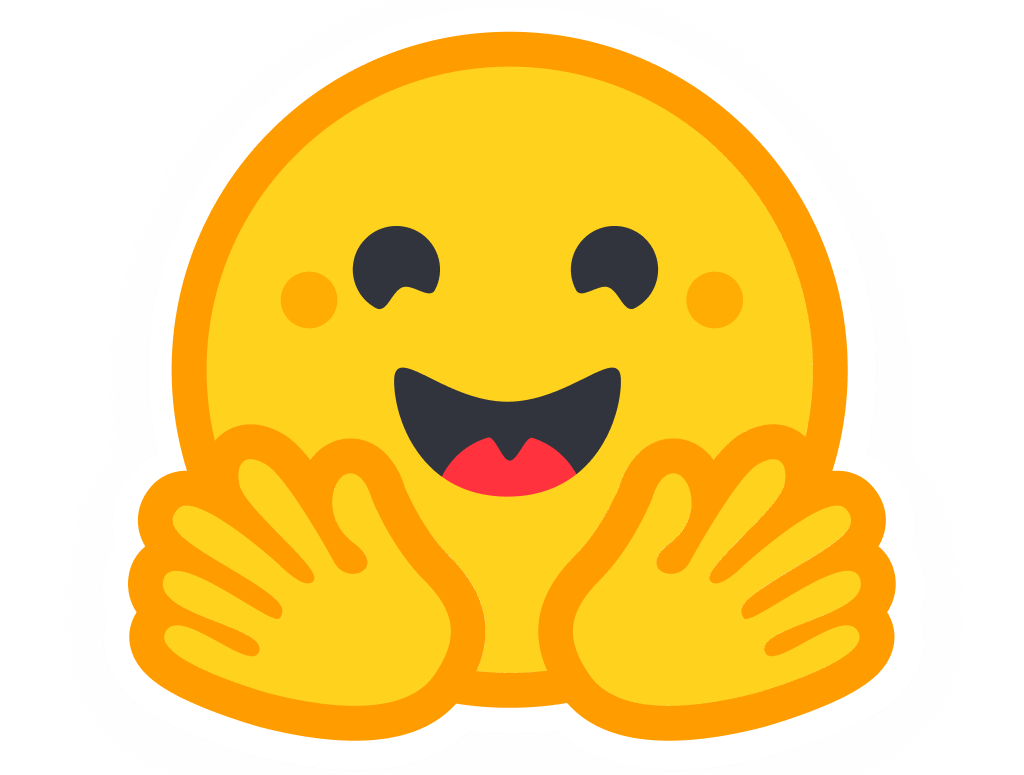}}}

\begin{document}
\maketitle
\hspace*{\fill} \\
\hspace*{\fill} \\
\begin{abstract}
Decomposition-based multi-hop retrieval methods rely on many autoregressive steps to break down complex queries, which breaks end-to-end differentiability and is computationally expensive. Decomposition-free methods tackle this, but current decomposition-free approaches struggle with longer multi-hop problems and generalization to out-of-distribution data. To address these challenges, we introduce \textbf{GRITHopper-7B}\footnotemark[1], a novel multi-hop dense retrieval model that achieves state-of-the-art performance on both in-distribution and out-of-distribution benchmarks. GRITHopper combines generative and representational instruction tuning by integrating causal language modeling with dense retrieval training. Through controlled studies, we find that incorporating additional context after the retrieval process, referred to as \emph{post-retrieval language modeling}, enhances dense retrieval performance. By including elements such as final answers during training, the model learns to better contextualize and retrieve relevant information. GRITHopper-7B offers a robust, scalable, and generalizable solution for multi-hop dense retrieval, and we release it to the community\footnotemark[2] for future research and applications requiring multi-hop reasoning and retrieval capabilities.

\footnotetext[1]{\href{https://huggingface.co/UKPLab/GritHopper-7B}{\huggingfacelogo\hspace{0.05cm} UKPLab/GritHopper-7B}}
\footnotetext[2]{\href{https://github.com/UKPLab/EACL2026-GritHopper}{\githublogo\hspace{0.1cm}UKPLab/EACL2026-GritHopper}}

%Our model and code will be published upon acceptance.
%Our model\footnotemark[1] and code\footnotemark[2] are publicly available.

%\footnotetext[2]{\href{TODO}{\githublogo\hspace{0.1cm}Our Repo}}
%\footnotetext[1]{\href{TODO}{\huggingfacelogo\hspace{0.05cm} GRITHopper-7B}}

% \item Interpretability results: show that entity guessing should be impossible from an information theory perspective (make analysis for type of questions --> match who, how, where questions and analyse the topics (topic similarity) --> say that this is a study of construction from 
%\item implement MDR experiments for Beam Retriever (See paper last page HotpotQA Open Retrival

\end{abstract}

\section{Introduction}

Large Language Models (LLMs) have demonstrated remarkable capabilities in reasoning \cite{huang-chang-2023-towards}, reflection, and decomposition, making them indispensable tools for a wide range of natural language processing tasks. Their generative abilities have been successfully leveraged to solve open-domain multi-hop problems, where complex questions are broken into smaller sub-questions to retrieve supporting evidence and reflect on them \citep{asai2024selfrag, shao-etal-2023-enhancing, guan2024amor} in a step-by-step manner. However, such decomposition-based approaches require multiple autoregressive steps and discrete intermediate outputs, which breaks the end-to-end differentiability of the retrieval pipeline and increases computational overhead.
\begin{figure}
    \centering
    \includegraphics[width=1.0\linewidth]{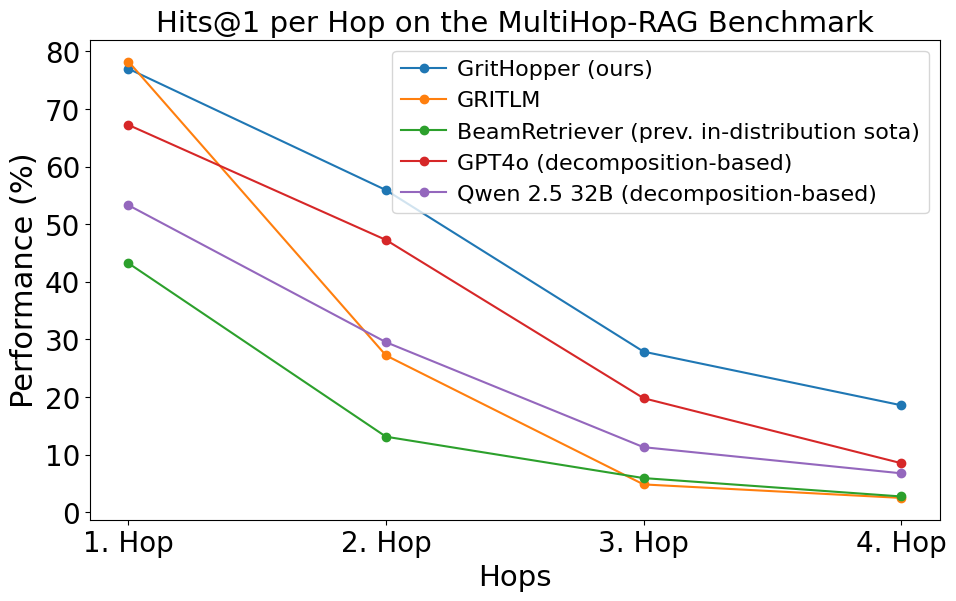}
    \caption{Out-of-distribution Multi-Hop Retrieval Performance on the MultiHop-RAG Benchmark \cite{tang2024multihoprag}. While methods such as BeamRetriever achieve state-of-the-art performance in in-distribution settings, GRITHopper consistently outperforms them under distribution shift, with the relative advantage increasing at deeper hops, highlighting its robustness to long-context accumulation.}
    \label{fig:1}
\end{figure}
\begin{figure*}
    \centering
    \includegraphics[width=1.0\linewidth]{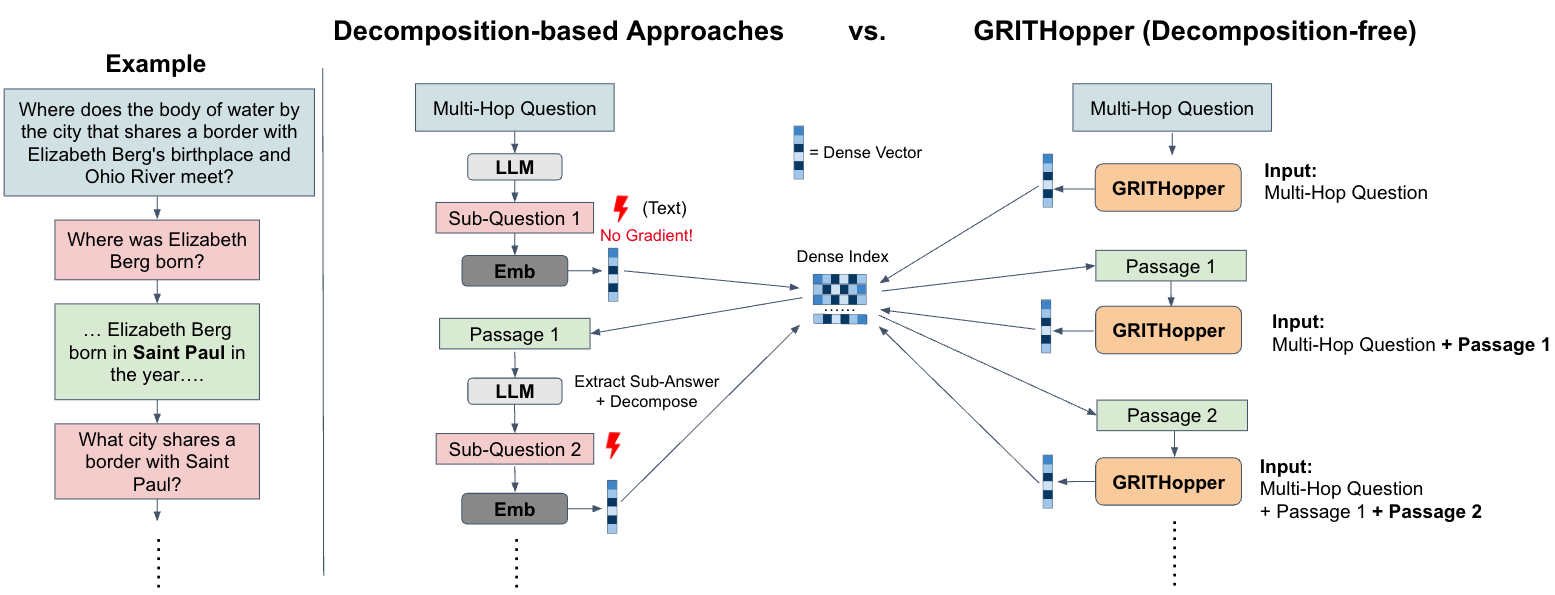}
    \caption{Comparison of decomposition-based approaches like \citep{guan2024amor,shao-etal-2023-enhancing} to our encoder-only approach with GRITHopper. While decomposition-based approaches require many auto-regressive steps to decompose questions, extract answers, and a different model for retrieval, our encoder-only approach only requires a single forward pass per hop to compute the next dense vector. Example is from \cite{trivedi-etal-2022-MuSiQue}.}
    \label{fig:2} 
\end{figure*}

Decomposition-free approaches, such as Multi-Hop Dense Retrieval (MDR) \cite{xiong2021answering}, and cross-encoder-based methods like Beam Retriever \cite{zhang-etal-2024-end}, enable end-to-end differentiability by not requiring discrete decompositions, but both suffer from significant limitations. MDR offers an efficient and scalable dense retrieval framework by concatenating the query with passages and encoding them into a single vector representation in one model call per iteration. However, it struggles with more complex datasets like MuSiQue \cite{trivedi-etal-2022-MuSiQue}, more hops than 2, and performs poorly on out-of-distribution benchmarks. On the other hand, Beam Retriever achieves state-of-the-art in-distribution performance by leveraging cross-encoder architectures. Unlike bi-encoders, which independently encode questions and passages to compute similarity, cross-encoders process both as a single sequence, resulting in linear scaling with respect to the number of passages. This makes them only suited as a retriever for a few hundred passages but not open book retrieval. Despite its strengths, it shares MDR’s generalization issues while introducing scalability challenges due to its computational overhead, making it impractical for large-scale open retrieval tasks. These limitations underscore the need for a scalable and generalizable multi-hop retrieval framework that can perform well on both in-distribution and out-of-distribution benchmarks in open-domain retrieval scenarios.

To address these challenges, we introduce GRITHopper-7B, the first decoder-based end-to-end multi-hop dense retrieval model trained on an unprecedented scale of multi-hop datasets spanning both question-answering and fact-checking tasks. GRITHopper-7B achieves state-of-the-art performance across out-of-distribution benchmarks (see Figure \ref{fig:1}) while preserving the simplicity and scalability of encoder-only paradigms like MDR (see Figure \ref{fig:2}). The foundation of GRITHopper lies in GRITLM \cite{muennighoff2024generative}, a Mistral-7B-based model that integrates causal language modeling with dense retrieval training. GRITLM’s design sparked a critical debate in the field: \textit{Does joint optimization of generative and retrieval tasks enhance dense embedding quality?} While GRITLM initially demonstrated state-of-the-art results in retrieval while achieving strong performance in generation, subsequent studies \cite{lee2025nvembed} show that contrastive-only approaches, using the same Mistral-7B backbone, outperform GRITLM on key benchmarks such as BEIR \cite{thakur2021beir} and MTEB \cite{muennighoff-etal-2023-mteb}. 

This raises fundamental questions about the utility of generative objectives in retrieval and sets the stage for a deeper exploration of their role. Building upon a shared data foundation for both the retrieval and generation objective, we incrementally add information to the generative component without altering the embedding component. This strategy allows us to assess whether incorporating external information (beyond the retrieval chain) into the generative training can improve dense retrieval performance. We refer to this approach as \textit{post-retrieval language modeling}, where we include elements such as final answers and judge the retrieved paragraphs after the retrieval chain. Through this controlled experimental setup, we systematically explore how post-retrieval language modeling influences dense embedding quality, offering new insights into their roles in enhancing multi-hop retrieval performance. Our experiments create a novel ReAct style \cite{yao2023react} end-to-end multi-hop dense retrieval that can conduct neural search via bi-directional attention and control itself (stop the search, answer, or rerank) via causal language modeling.

The following research questions guide our study:

\noindent\textbf{RQ1}: What is the effect of combining generative and embedding training in multi-hop dense retrieval compared to embedding-only training?

\noindent\textbf{RQ2}: If generative training improves dense retrieval performance, does post-retrieval language modeling during training further enhance it?

\noindent\textbf{RQ3}: How does GRITHopper generalize on the out-of-distribution benchmarks compared to existing methods?

\noindent\textbf{RQ4}: How do decomposition-free approaches compare to decomposition-based approaches?

\section{Related Work}\label{sec:related}

\subsection{Multi-Hop Retrieval and Reasoning}

Multi-hop question answering requires models to retrieve and integrate information from multiple documents to answer complex queries  \citep{trivedi-etal-2022-MuSiQue,ho-etal-2020-constructing}. Decomposition-based methods address this by breaking down complex questions into simpler sub-questions. \citet{wolfson-etal-2020-break} introduced the Break It Down (Break) method, which decomposes questions into a sequence of simpler queries. Other methods extended decompositions with extensive reasoning \citep{shao-etal-2023-enhancing,khot2023decomposed,yao2023react}. However, these methods require multiple autoregressive steps and generate intermediate outputs, leading to increased computational overhead and disrupting end-to-end differentiability. Decomposition-free approaches have been proposed to overcome these limitations.

\subsection{Decomposition-Free Multi-Hop Retrieval}
Multi-Hop Dense Retrieval (MDR) \cite{xiong2021answering} introduced an approach where the query is concatenated with previously retrieved passages, and the combined text is encoded into a single vector representation using a bi-encoder architecture. Other works have extended MDR, such as BeamDR by adding beam search and \citet{ma-etal-2024-ex} by extending MDR multi-hop problems longer than 2 hops. While MDR allows for efficient and scalable retrieval but has limitations in handling complex multi-hop queries that require more hops than 2 and generalizing to unseen datasets.

Multi-Hop cross-encoder models \cite{asai2020learning}, like the BeamRetriever \cite{zhang-etal-2024-end}, achieve state-of-the-art performance on in-distribution datasets by modeling the retrieval process by encoding the question with each paragraph together. Despite their effectiveness, these models face scalability issues due to high computational costs, making them less practical for large-scale open-domain retrieval tasks. Furthermore, we will show that these methods suffer from overfitting and fail to generalize on out-of-distribution benchmarks.

\subsection{Causal Language Modeling and Reward Modeling} % \cite{radford2019language}
 While Causal language modeling (CLM) is primarily used for generation tasks \cite{radford2019language}, recent research has combined it with dense retrieval, specifically GRITLM \citet{muennighoff2024generative}, integrating causal language modeling with contrastive learning by simply adding the next token and contrastive loss. While the method trained on two distinct datasets for retrieval and generation, it leaves much room for exploration on how these two losses work together.

In language models, reward modeling can guide the generation process towards more accurate or contextually appropriate responses. \citet{zelikman2022star} and \citet{huang-chang-2023-towards} explored how self-taught reasoning and reflection can improve reasoning capabilities in language models, which could be beneficial for retrieval tasks that require complex reasoning. To distinguish positive from negative passages, we adopt the approach from \cite{zhang2024generative} that has shown that language models can simulate reward learning through simple next-token prediction. This comes especially handy for GRITLM's joint generative and embedding objective.

\section{Problem Statement \& Evaluation}
\subsection{Problem Definition}
In the context of multi-hop retrieval, given a fixed corpus of paragraphs P and a multi-hop-question \textit{q}, the task is to identify a sequence of paragraphs $[p_1,p_2...,p_n]$ where $p_i \in P$, that collectively answer $q$ \citep{trivedi-etal-2022-MuSiQue, ho-etal-2020-constructing}. 
Decomposition-free methods \citep{xiong2021answering, zhang-etal-2024-end} concatenate the multi-hop question together with previously retrieved paragraphs  $[q, p_1, p_2, .., p_n]$ on the word level and feed them as a single string into an Encoder model E to retrieve the next paragraph as:
\begin{equation}
E(q, p_1, p_2, \dots, p_n) \rightarrow E(p_{n+1})
\end{equation}

where all candidate passages $p_{n+1} \in P$ are pre-computed offline. Apart from question answering, we also adapt fact-checking retrieval as $[claim, p_1, p_2, .., p_n]$ where paragraphs can either be supporting or refuting paragraphs.

\subsection{Datasets}\label{sec:data}
We train all models on MuSiQue \cite{trivedi-etal-2022-MuSiQue}, HotpotQA \cite{yang-etal-2018-HotpotQA}, 2WikiMultiHopQA \cite{ho-etal-2020-constructing}, Explainable Fever \cite{ma-etal-2024-ex}, and HoVer \cite{jiang-etal-2020-hover}. These datasets encompass question-answering and fact-checking tasks with varying levels of complexity and hop depths. For out-of-distribution evaluation, we use the MultiHopRAG Benchmark \cite{tang2024multihoprag} and MoreHopQA \cite{schnitzler2024morehopqamultihopreasoning}.

\subsection{Evaluation}
To demonstrate the performance of all approaches at different hop depths, we calculate Hits@k at each hop. This metric considers a hop successful if the relevant passage is retrieved within the top-k results. Importantly, the evaluation only continues to the next hop if the previous hop was successful. This allows us to analyze the performance across varying hop depths, highlighting the ability of models to retrieve relevant passages in a sequential multi-hop setup. If not explicitly mentioned, we evaluate only the dense retrieval performance. In our end-to-end evaluation, we also measure the performance of the model to decide when to stop retrieval.

\section{Methods}\label{sec:methods}
Our central objective is to understand how integrating causal language modeling (CLM) with dense embedding training impacts multi-hop retrieval \textbf{(RQ1)}, and whether adding post-retrieval signals (e.g., final answers, judging hard negatives) can further improve performance \textbf{(RQ2)}. Unlike prior work, \cite{muennighoff2024generative}, which combined generative and embedding training on \emph{different} datasets, we investigate their interplay under a unified, controlled setup. This allows us to isolate the influence of the generative objective on embedding quality. Previous research in language model pretraining has shown that combining masked language modeling (MLM) with embedding training on the \emph{same} dataset often improves downstream representations \cite{devlin-etal-2019-bert,wu-etal-2020-tod}. 

\begin{figure}
    \centering
    \includegraphics[width=1.0\linewidth]{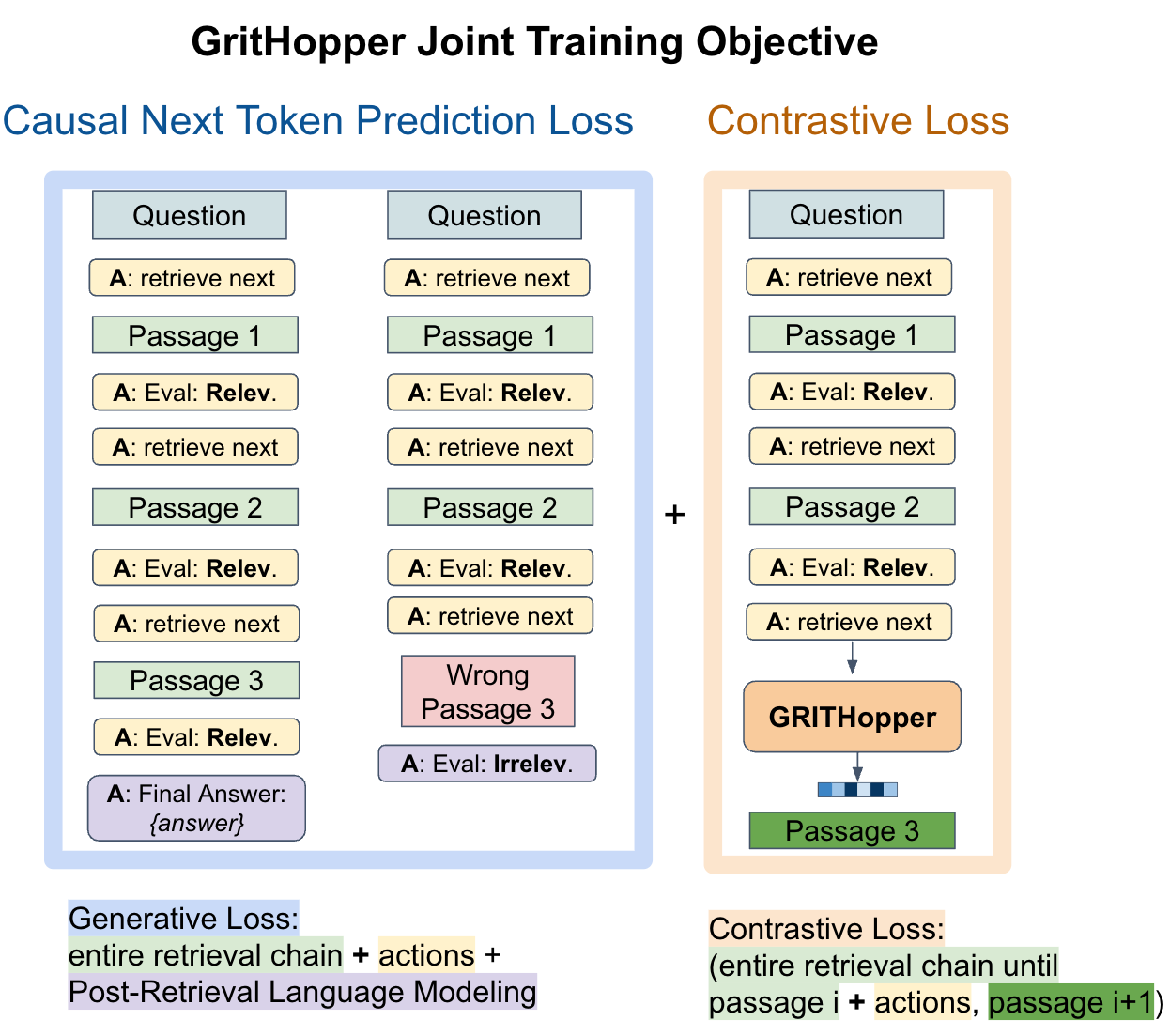}
    \caption{Highlighting the joint training objective (generative and contrastive) of GRITHopper. Both objectives consume the exact same tokens, except for the post-retrieval added information to the generative loss in purple. Note that if the model is used like MDR without a stopping condition, we keep one forward pass per hop to generate the embedding, as all action tokens are only prompt tokens (not output tokens). Only if we want to use the framework end-to-end by controlling when to stop/conduct reranking do we have to do one/two additional causal forward passes.}
    \label{fig:3}
\end{figure}

\subsection{A Shared Dataset for a Controlled Setup}
To critically evaluate how CLM and embedding objectives affect each other, we start from a \emph{shared dataset}, where both objectives consume identical tokens. Concretely, consider a multi-hop question $q$ and the sequence of previously retrieved paragraphs $[p_1, p_2, \dots, p_n]$. The embedding model learns to represent $[q, p_1, \dots, p_n]$ so that it can retrieve the next relevant paragraph $p_{n+1}$, while the generative model predicts the next tokens on the same sequence in a causal manner. This controlled baseline ensures that any retrieval improvement upon adding the generative loss cannot be attributed to extraneous factors like domain shifts or additional training data. Instead, it must arise from the generative objective itself, addressing \textbf{RQ1}: does integrating CLM with embedding training, under controlled conditions, enhance retrieval?

Starting from this shared dataset, we then incrementally enrich the generative model’s input with post-retrieval information while keeping the embedding input fixed. This step-by-step strategy ensures that each addition’s impact on retrieval is transparent and attributable solely to the newly introduced elements, addressing \textbf{RQ2}. To make this process explicit, we experiment with two types of post-retrieval information we add to the generative input while keeping the retrieval-side embeddings fixed: (1) appending the final answer to ensure the model predicts it conditioned on retrieved context, and (2) introducing irrelevant passages as causal negatives to teach discrimination during generation.

\noindent\textbf{1. Adding Final Answers:}  
We append the final answer $ans$ to the retrieval chain:
\[
[q, p_1, p_2, \dots, p_n, ans].
\]
The embedding objective gets the exact same tokens $[q, p_1, \dots, p_n]$ as the generative objective, while the generative objective now additionally predicts the $ans$.

\noindent\textbf{2. Adding Hard Negatives:}  
We further augment the generative training by introducing an irrelevant passage $p_{ir}$, marked as such:
\[
[q, p_1, p_2, \dots, p_{n-1}, p_{ir}, \text{Irrelevant}].
\]

The model learns to label the irrelevant paragraph causally via next-token prediction \cite{zhang2024generative}. If retrieval benefits from this, it indicates that contrasting positive and negative evidence in a generative framework helps refine the embedding space. This incremental approach, starting from a pure shared dataset and progressively adding final answers and negatives, provides a precise experimental lens. We directly measure how each augmentation in the generative domain influences the embedding model’s retrieval capabilities.

\subsection{ReAct-Style Instruction Tuning for End-to-End Multi-Hop Retrieval}
To incorporate these different actions to represent the entire multi-hop retrieval as a coherent textual sequence, we adapt the ReAct framework \cite{yao2023react}. Each retrieval hop, document evaluation, and final answer production is expressed as a short instruction or “action” phrase (see Figure \ref{fig:3}).

All these actions are represented as textual strings and integrated into the same sequences used by both the embedding and generative objectives. Their exact formatting for all multi-hop datasets (see §\ref{sec:data}) is described in Algorithm~\ref{alg:dataset-construction} in Appendix \ref{sec:algo}. Because these augmented sequences include both the retrieval chain (i.e., $[q, p_1, \dots]$) and the action strings, we maintain the shared data dataset principle for both embedding and generative training. This ReAct adaptation allows us to combine everything, final answers, negative passages, and retrieval steps, into a single, end-to-end system. Crucially, this framework allows the model to:
\begin{itemize}[noitemsep,nolistsep]
    \item Decide if a retrieved document is relevant or not. (\texttt{Eval} in Figure \ref{fig:3})
     \item Stop the search early if it encounters an irrelevant paragraph. (after (\texttt{Eval: Irrelev.}) in Figure \ref{fig:3})
      \item Continue retrieving until all information is gathered (\texttt{retrieve next} in Figure \ref{fig:3})
       \item Produce the answer. (\texttt{Final Answer:} in Figure \ref{fig:3})
\end{itemize}

In other words, the ReAct-style instruction tuning not only aligns with our controlled experimental design but also yields a system capable of autonomously handling the retrieval pipeline end-to-end. The model can determine how many steps to take and when to stop while providing a realistic and comprehensive testbed for studying the interplay of CLM and embedding objectives in multi-hop retrieval.

\section{Experimental Setup}\label{sec:expSetup}
% TODO add the number of samples
We train GRITHopper in two different setups. First, we explore our ablations by fine-tuning one dataset, MuSiQue \cite{trivedi-etal-2022-MuSiQue}. MuSiQue offers decomposition steps with which we can ensure highly qualitative hard negatives and is the most difficult multi-hop question answering dataset in our dataset collection, according to \citet{trivedi-etal-2022-MuSiQue}. Furthermore, we train our core ablations on a large collection of multi-hop datasets (described in §\ref{sec:data}) on two seeds. We explore in Appendix \ref{sec:detailedData} how we adapt each dataset in detail and describe the hard negative mining in Appendix \ref{sec:neg}. We discuss the training setup in Appendix \ref{app:trainingSetup}.

\subsection{Baselines} \label{sec:base}
Our baselines can be split into decomposition-free approaches and decomposition-based approaches. Starting with decomposition-free approaches, we chose GRITLM as our first baseline with the prompting formats we utilize for GRITHopper. GRITLM has also been trained on multi-hop question answering on HotpotQA and several Fever datasets \cite{thorne-etal-2018-fever} for single-step retrieval. Secondly, we train BeamRetriever (beam size 1), the current state-of-the-art method for multi-hop retrieval and MDR, on MuSiQue as well as our entire dataset collection (see §\ref{sec:data}). However, MDR has only been trained on a fixed number of 2 hops. Therefore, we remove any additional hops after the second hop in our experiments. For MDR, we choose RoBerta-Large \cite{liu2019robertarobustlyoptimizedbert}, and for BeamRetriever and Deberta-v3-Base \cite{he2021debertav3}, we find that these models perform best among Large and XL variations with the corresponding architectures. For more details on how we explored different base models for these architectures, see appendix \ref{app:base}. Besides decomposition-free methods like GRITHopper, BeamRetriever, and MDR, we add an additional baseline using decompositions. For this, we employ a simple one-step-at-a-time decomposition (like \cite{guan2024amor} but with only one try for a fair comparison) method using Qwen 2.5 32B (and GPT4o on two datasets) for decomposing the multi-hop problem into a single sub-question with 4 few-shot samples. The four few-shot examples used for sub-question generation were randomly sampled from the training set, where at hop h we select only few-shot examples taken from hop h of multi-hop training instances. In the second step, we use GRITLM to embed the sub-query and retrieve candidates. If a supporting paragraph is retrieved within the top-k range, we continue by asking Qwen/GPT4o to extract the answer and use the previously solved sub-questions to decompose the next sub-query. We provide the prompt templates and GPT4o generation outputs in the appendix \ref{sec:decomp}. 
\section{Experiments and Discussion}\label{sec:experiments}
\label{sec:eval}

In this section, we first investigate GRITHopper’s ablations in detail, as these represent the core baselines for the first decoder-based multi-hop dense retrieval model. We then compare GRITHopper to existing methods in an open retrieval setting, including decomposition-free BERT-based models (MDR, BeamRetriever), general instruction-tuned retrieval models (GRITLM), and decomposition-based approaches with GPT-4o and Qwen. Subsequently, we analyze GRITHopper’s out-of-distribution generalization capabilities (RQ3) and focus specifically on decomposition-based methods (RQ4), highlighting its robustness over previous state-of-the-art approaches. We discuss inference compute in Appendix~\ref{sec:complexity_appendix} and training time in Appendix~\ref{sec:trainingtime}.
\begin{table}[t]
    \centering
    \small
    \begin{tabular}{l c}
        \toprule
        \textbf{Model} & \textbf{Average Hits@1} \\
        \midrule
        \multicolumn{2}{l}{\textbf{Dense Retrieval}} \\
        \midrule
        GRITHopper (Answers \& Negatives)  & \textbf{82.32} \\
        GRITHopper (Answers)            & 82.08 \\
        GRITHopper (no post lm)         & 80.78 \\
        GRITHopper (Contrastive Only)   & 78.02 \\
        \midrule
        \multicolumn{2}{l}{\textbf{Cross Encoder}} \\
        \midrule
        BeamRetriever Large (all datasets)    & \textbf{85.10} \\
        BeamRetriever (all datasets)    & 81.78 \\
        BeamRetriever (MuSiQue Only)    & 80.98 \\
        GRITHopper ReRank$^*$           & 59.04 \\
        \midrule
        \multicolumn{2}{l}{\textbf{End-to-End Retrieval}} \\
        \midrule
        GRITHopper end-to-end$^*$       & \textbf{75.00} \\
        BeamRetriever end-to-end  & 38.21 \\
        \bottomrule
    \end{tabular}
\caption{MuSiQue distractor-setting dense retrieval performance. All GRITHopper models are trained only on the MuSiQue dataset. $^*$ Uses GRITHopper (Answers \& Negatives). \texttt{No post lm} stands for causal modeling only on the retrieval chain. For end-to-end retrieval, differences in performance are closely tied to stopping behavior and per-hop latency; a detailed analysis of early stopping, overshooting, and latency is provided in Appendix~\ref{app:diagnostics}.}
    \label{tab:distractor}
\end{table}

\begin{table}[t]
    \centering
    \scriptsize
    \begin{tabular}{l c c c}
        \toprule
        \textbf{Dataset} & \multicolumn{3}{c}{\textbf{Avg. Hits@1 for GRITHopper with:}} \\
        \cmidrule(lr){2-4}
                         & \textbf{Ans + Neg} & \textbf{Ans} & \textbf{No Post} \\
        \midrule
        \textbf{In Distribution} & & & \\
        ExFever                 & 87.10           & \textbf{91.81} & 89.69 \\
        MuSiQue                 & \textbf{76.16}           & 75.95 & 75.22 \\
        Hover                   & 93.34          & \textbf{94.29} & 94.36 \\
        \midrule
        \textbf{Zero-Shot Benchmarks} & & & \\
        MoreHopQA               & \textbf{96.14}           & 95.80 & 94.68 \\
        MultiHopBench           & 51.74  & \textbf{54.03}          & 51.13 \\
        \bottomrule
    \end{tabular}
    \caption{GRITHopper trained on \textbf{all datasets} in open retrieval performance. Results are averaged over two seeds. \textbf{Ans} includes the final answer in the generative samples. \textbf{Neg} includes reward modeling on negatives while \textbf{No Post} does not include post-retrieval language modeling.}
    \label{tab:my_label}
\end{table}

\subsection{Evaluating Generative Objectives and Post-Retrieval Information (RQ1, RQ2)}
As GritHopper is the first decoder-based decomposition-free Multi-Hop Dense retriever, we extensively ablate our training objective to motivate our auxiliary training signals, including

\begin{enumerate}[noitemsep]
    \item only contrastive learning (like MDR, just GRITLM fine-tuned on multi-hop datasets)
    \item contrastive + causal language modeling with no post-retrieval information (same data for causal + contrastive)
    \item contrastive + causal language modeling with final answers
    \item contrastive + causal language modeling with final answers and causal negative
\end{enumerate}
to address our research questions, RQ1 \& RQ2.

We first conduct a series of controlled experiments on the MuSiQue dataset under the distractor setting (around 20 handcrafted distractor documents) (see Table~\ref{tab:distractor}) and then move to training on all datasets in Table~\ref{tab:my_label} on open-retrieval averaged across two seeds. This scenario allows us to isolate and compare the effects of different generative strategies (with and without final answers) and reward modeling on negative documents before deploying the chosen configurations in the more challenging open retrieval environment.

On MuSiQue, our best-performing GRITHopper variant uses both final answers and negative documents, achieving a Hits@1 score of 82.32. Even without reward modeling on negatives, adding the final answer results in a still-impressive Hits@1 score of 82.08. Compared to a purely contrastive approach (like MDR) without generative signals (78.02), these findings demonstrate that causal language modeling on the same dataset (80.78) improves performance (RQ1). Building on that, the inclusion of final answers (part of RQ2) substantially improves retrieval accuracy (82.08) and is essential for outperforming BeamRetriever in distribution on MuSiQue (81.78). The final answer during training provides a clearer retrieval target, guiding the model to select more relevant passages at each hop. 
\subsubsection*{Open Retrieval}
However, when scaling these ablations to all datasets (Table~\ref{tab:my_label}), reward modeling, while effective in the distractor setting, led to overfitting in open retrieval. Specifically, the GRITHopper observing negatives causally during training (cross-encoder training) caused a 7.32\% drop in Average Hits@1 when transitioning from the distractor setting to open retrieval on MuSiQue, compared to a milder 5.09\% drop for its counterpart (only with Answers), averaged over two seeds. This is even more extreme with BeamRetriever, which excels under conditions closely matching its training distribution (distractor setting in Table~\ref{tab:distractor}) but struggles to generalize on the same dataset in open retrieval (Table~\ref{tab:your_label}). Here, the DeBerta Large version, while achieving the strongest results under distractors (see Table~\ref{tab:distractor}), performs worse than the base variant in open retrieval; we explore this further in Appendix~\ref{app:beam}. These findings suggest that learning difficult negatives causally can improve discrimination on difficult distractors but hinder broader generalization in dense retrieval. By contrast, GradCache’s large in-batch negatives provide a more robust discriminative learning signal while having a slight disadvantage of “hand-crafted” distractor discrimination. Thus, while both generative training and final answers prove beneficial (answering RQ1 and partially RQ2 affirmatively), reward modeling offers only limited gains and at a considerable cost to generalization. 
\subsubsection*{End-to-End Retrieval}
Furthermore, we compare the end-to-end performance of the models in terms of stopping after the correct number of hops. BeamRetriever determines when to stop by comparing the retrieval score of the current hop to that of the previous hop and terminates once the score decreases (see \cite{zhang-etal-2024-end}, Appendix C). However, we find that these scores are biased to decrease after the first hop, often leading to premature stopping. GRITHopper appears more robust in this scenario (see Table~\ref{tab:distractor}). We provide a more detailed analysis of stopping behavior and per-hop latency in Appendix~\ref{app:diagnostics}, showing that GRITHopper achieves substantially lower latency per hop due to its dense retrieval design. We further observe a slight misalignment between causal and dense retrieval performance, which we explore in Appendix~\ref{app:causvsdense}. Importantly, we emphasize that all reported checkpoints are selected based on peak dense retrieval performance rather than generative accuracy, which explains why GRITHopper’s generative behavior may appear weaker compared to methods optimized explicitly for generation.

%However, we find a slight misalignment in the causal and dense retrieval performance, which we explore in Appendix \ref{app:causvsdense}.

\subsection{Comparison to Existing Methods on Open Retrieval}
\begin{table*}[h!]
\centering
\footnotesize % Further reduces font size for better spacing
\renewcommand{\arraystretch}{1.1} % Increases row height for readability
\setlength{\tabcolsep}{4pt} % Slightly increases column spacing
\definecolor{lightgray}{gray}{0.9} % Define a light gray color
\resizebox{\textwidth}{!}{%
\begin{tabular}{l|ccccc|ccccc|ccccc|}
\toprule
  \multirow{2}{*}{Model} & \multicolumn{5}{c|}{Hits@1} & \multicolumn{5}{c|}{Hits@5} & \multicolumn{5}{c|}{Hits@10} \\
  \cmidrule(lr){2-6} \cmidrule(lr){7-11} \cmidrule(lr){12-16}
   & 1 & 2 & 3 & 4 & \cellcolor{lightgray}\itshape Avg & 1 & 2 & 3 & 4 & \cellcolor{lightgray}\itshape Avg & 1 & 2 & 3 & 4 & \cellcolor{lightgray}\itshape Avg \\
\midrule

\multicolumn{16}{l}{\textbf{MuSiQue}} \\
GRITHopper (ours) & 94.25 & 76.13 & 55.45 & 32.10 & \cellcolor{lightgray}\itshape 76.42 & 99.59 & 96.32 & 85.92 & 57.04 & \cellcolor{lightgray}\itshape 93.18 & 99.79 & 98.59 & 91.07 & 69.63 & \cellcolor{lightgray}\itshape 95.85 \\
GRITLM  & 91.15 & 57.51 & 22.32 & 5.43 & \cellcolor{lightgray}\itshape 60.51 & 99.50 & 91.31 & 65.49 & 35.56 & \cellcolor{lightgray}\itshape 86.18 & 99.96 & 96.61 & 83.26 & 51.85 & \cellcolor{lightgray}\itshape 92.61 \\
MDR & 81.75 & 45.18 & - & - & \cellcolor{lightgray}\itshape 63.47 & 94.37 & 71.04 & - & - & \cellcolor{lightgray}\itshape 82.71 & 96.73 & 78.82 & - & - & \cellcolor{lightgray}\itshape 87.77 \\
Beam Retriever  & 88.75 & 60.70 & 30.73 & 12.84 & \cellcolor{lightgray}\itshape 62.80 & 95.45 & 85.40 & 65.84 & 41.48 & \cellcolor{lightgray}\itshape 82.85 & 97.02 & 90.44 & 77.25 & 51.60 & \cellcolor{lightgray}\itshape 88.07 \\
%Beam Retriever Large & 82.79 & 59.54 & 35.02 & 15.80 & \cellcolor{lightgray}\itshape 61.09 & 91.73 & 85.35 & 72.36 & 44.69 & \cellcolor{lightgray}\itshape 82.82 & 95.04 & 92.43 & 84.55 & 57.78 & \cellcolor{lightgray}\itshape 89.79 \\
Qwen 2.5 32B + GRITLM decomposition & 82.62 & 45.72 & 13.91 & 1.48 & \cellcolor{lightgray}\itshape 51.06 & 95.45 & 76.25 & 36.05 & 13.09 & \cellcolor{lightgray}\itshape 72.19 & 96.69 & 82.91 & 46.61 & 17.78 & \cellcolor{lightgray}\itshape 77.39 \\
\textbf{GPT4o} + GRITLM decomposition & 81.96 & 48.53 & 13.39 & 1.98 & \cellcolor{lightgray}\itshape 51.81 & 95.82 & 79.19 & 33.39 & 9.63 & \cellcolor{lightgray}\itshape 72.74 & 97.35 & 85.35 & 42.23 & 14.81 & \cellcolor{lightgray}\itshape 77.58 \\

\midrule

\multicolumn{16}{l}{\textbf{Explainable Fever}} \\
GRITHopper (ours) & 96.88 & 92.20 & 85.38 & - & \cellcolor{lightgray}\itshape 93.02 & 99.79 & 99.29 & 98.72 & - & \cellcolor{lightgray}\itshape 99.40 & 99.94 & 99.53 & 99.13 & - & \cellcolor{lightgray}\itshape 99.63 \\

GRITLM   & 91.13 & 54.88 & 17.28 & - & \cellcolor{lightgray}\itshape 63.83 & 99.47 & 82.89 & 41.89 & - & \cellcolor{lightgray}\itshape 82.99 & 99.79 & 88.47 & 51.98 & - & \cellcolor{lightgray}\itshape 87.12 \\
MDR & 92.93 & 77.16 & - & - & \cellcolor{lightgray}\itshape 85.13 & 99.08 & 94.11 & - & - & \cellcolor{lightgray}\itshape 96.62 & 99.44 & 95.97 & - & - & \cellcolor{lightgray}\itshape 97.72 \\
Qwen 32B + GRITLM decomposition  & 63.24 & 29.88 & 11.93 & - & \cellcolor{lightgray}\itshape 40.90 & 83.74 & 55.14 & 31.87 & - & \cellcolor{lightgray}\itshape 63.27 & 88.96 & 63.61 & 40.14 & - & \cellcolor{lightgray}\itshape 70.34 \\

\midrule

\multicolumn{16}{l}{\textbf{HoVer}} \\
GRITHopper (ours) & 95.86 & 91.56 & 91.69 & 92.31 & \cellcolor{lightgray}\itshape 93.88 & 99.79 & 99.61 & 99.43 & 100.00 & \cellcolor{lightgray}\itshape 99.69 & 99.95 & 99.68 & 99.71 & 100.00 & \cellcolor{lightgray}\itshape 99.83 \\

GRITLM  & 95.81 & 88.09 & 83.95 & 88.46 & \cellcolor{lightgray}\itshape 91.81 & 99.89 & 99.53 & 98.28 & 96.15 & \cellcolor{lightgray}\itshape 99.57 & 99.89 & 99.76 & 98.85 & 100.00 & \cellcolor{lightgray}\itshape 99.74 \\
MDR & 84.77 & 65.69 & - & - & \cellcolor{lightgray}\itshape 77.10 & 96.60 & 89.51 & - & - & \cellcolor{lightgray}\itshape 93.75 & 97.98 & 92.51 & - & - & \cellcolor{lightgray}\itshape 95.78 \\
Beam Retriever & 98.04 & 88.96 & 85.96 & 76.92 & \cellcolor{lightgray}\itshape 93.42 & 99.47 & 97.56 & 97.71 & 100.00 & \cellcolor{lightgray}\itshape 98.61 & 99.73 & 97.79 & 97.71 & 100.00 & \cellcolor{lightgray}\itshape 98.84 \\
Qwen 32B + GRITLM decomposition & 75.38 & 61.44 & 50.43 & 46.15 & \cellcolor{lightgray}\itshape 67.69 & 82.23 & 74.84 & 68.19 & 69.23 & \cellcolor{lightgray}\itshape 78.09 & 84.24 & 78.15 & 72.21 & 73.08 & \cellcolor{lightgray}\itshape 80.78 \\

\midrule

\multicolumn{16}{l}{\textbf{Zero-Shot Multi-Hop RAG Benchmark}} \\
GRITHopper (ours) & 76.98 & 55.92 & 27.89 & 18.59 & \cellcolor{lightgray}\itshape 55.87 & 98.63 & 89.22 & 60.97 & 51.76 & \cellcolor{lightgray}\itshape 84.80 & 99.78 & 94.90 & 71.43 & 64.32 & \cellcolor{lightgray}\itshape 90.17 \\

GRITLM & 78.23 & 27.23 & 4.85 & 2.51 & \cellcolor{lightgray}\itshape 40.19 & 98.49 & 75.21 & 33.76 & 16.33 & \cellcolor{lightgray}\itshape 71.98 & 99.87 & 91.04 & 59.86 & 36.93 & \cellcolor{lightgray}\itshape 84.75 \\
MDR & 19.56 & 2.22 & - & - & \cellcolor{lightgray}\itshape 10.89 & 41.60 & 9.36 & - & - & \cellcolor{lightgray}\itshape 25.48 & 50.55 & 15.12 & - & - & \cellcolor{lightgray}\itshape 32.84 \\
Beam Retriever & 43.24 & 13.13 & 5.95 & 2.76 & \cellcolor{lightgray}\itshape 22.22 & 60.09 & 28.47 & 19.56 & 14.07 & \cellcolor{lightgray}\itshape 37.52 & 68.56 & 37.03 & 27.89 & 19.85 & \cellcolor{lightgray}\itshape 45.83 \\
%Beam Retriever  Large & 60.35 & 24.61 & 7.23 & 1.01 & \cellcolor{lightgray}\itshape 32.96 & 86.08 & 49.67 & 25.43 & 11.31 & \cellcolor{lightgray}\itshape 55.97 & 91.80 & 62.66 & 36.73 & 19.85 & \cellcolor{lightgray}\itshape 65.65 \\
%BeamRetriever ModernBert  Large & 62.97 & 18.18 & 3.49 & 0.75 & \cellcolor{lightgray}\itshape 30.80 & 86.78 & 46.30 & 16.58 & 6.53 & \cellcolor{lightgray}\itshape 52.96 & 92.55 & 61.29 & 29.25 & 14.57 & \cellcolor{lightgray}\itshape 63.63 \\
Qwen 32B + GRITLM decomposition & 53.30 & 29.53 & 11.31 & 6.78 & \cellcolor{lightgray}\itshape 33.33 & 79.56 & 60.27 & 36.05 & 28.89 & \cellcolor{lightgray}\itshape 60.68 & 86.74 & 71.00 & 50.09 & 42.96 & \cellcolor{lightgray}\itshape 70.96 \\
\textbf{GPT4o} + GRITLM decomposition & 67.23 & 47.27 & 19.81 & 8.54 & \cellcolor{lightgray}\itshape 46.83 & 91.18 & 79.51 & 49.91 & 29.15 & \cellcolor{lightgray}\itshape 74.82 & 96.41 & 88.12 & 64.80 & 47.74 & \cellcolor{lightgray}\itshape 84.04 \\
\midrule

\multicolumn{16}{l}{\textbf{Zero-Shot MoreHopQA}} \\
GRITHopper (ours) & 96.96 & 93.92 & - & - & \cellcolor{lightgray}\itshape 95.44 & 99.91 & 99.19 & - & - & \cellcolor{lightgray}\itshape 99.55 & 100.00 & 99.73 & - & - & \cellcolor{lightgray}\itshape 99.87 \\

GRITLM & 98.75 & 95.53 & - & - & \cellcolor{lightgray}\itshape 97.14 & 100.00 & 98.84 & - & - & 
\cellcolor{lightgray}\itshape 99.42 & 100.00 & 99.73 & - & - & \cellcolor{lightgray}\itshape 99.87 \\
MDR & 88.73 & 75.58 & - & - & \cellcolor{lightgray}\itshape 82.16 & 98.30 & 90.79 & - & - & \cellcolor{lightgray}\itshape 94.54 & 99.46 & 93.47 & - & - & \cellcolor{lightgray}\itshape 96.47 \\

Beam Retriever & 97.85 & 93.02 & - & - & \cellcolor{lightgray}\itshape 95.44 & 99.82 & 98.21 & - & - & \cellcolor{lightgray}\itshape 99.02 & 100.00 & 98.39 & - & - & \cellcolor{lightgray}\itshape 99.19 \\
%Beam Retriever Large & 88.73 & 75.58 & - & - & \cellcolor{lightgray}\itshape 82.16 & 98.30 & 90.79 & - & - & \cellcolor{lightgray}\itshape 94.54 & 99.46 & 93.47 & - & - & \cellcolor{lightgray}\itshape 96.47 \\

Qwen 32B + GRITLM decomposition & 96.24 & 55.19 & - & - & \cellcolor{lightgray}\itshape 75.72 & 99.55 & 65.38 & - & - & \cellcolor{lightgray}\itshape 82.47 & 100.00 & 68.78 & - & - & \cellcolor{lightgray}\itshape 84.39 \\

\midrule

\end{tabular}%
}
\caption{Open Retrieval comparison  on different hop depths. We compare our best GRITHopper (with Answers but no reward modeling) to BeamRetriever, GRITLM, MDR, and a decomposition-based approach.}
\label{tab:your_label}
\end{table*}

Table~\ref{tab:your_label} summarizes the performance of various models on both in-distribution and out-of-distribution benchmarks (RQ3) across different hop depths. We compare GRITHopper to GRITLM, BeamRetriever, MDR, and Qwen 32B / GPT4o + GRITLM with decompositions. While BeamRetriever represents the state of the art for in-distribution multi-hop retrieval, its performance degrades substantially under out-of-distribution evaluation, as reflected in our zero-shot benchmarks.

Across all evaluated tasks, GRITHopper consistently outperforms all other techniques, including the state-of-the-art model Beam-Retriever, while being significantly more efficient, as we explore in appendix \ref{app:speed}. For example, on the most difficult dataset, the out-of-distribution MultiHopRAG benchmark, GRITHopper, achieves a significant improvement in Hits@1 at deeper hops. GRITLM, a previous generative-retrieval hybrid model, performs well for the first hop but struggles with deeper hops. BeamRetriever, despite demonstrating strong performance in in-distribution tasks, exhibits a substantial performance drop when tested on the out-of-distribution MultiHopRAG benchmark, highlighting its tendency to overfit on training datasets. In contrast, GRITHopper maintains strong retrieval quality even when encountering unseen data (RQ3). MDR degrades the most in this scenario.

\subsection{Decomposition-Based Approaches (RQ4)}

We now turn our focus on the comparison to decomposition-based methods (RQ4). The Qwen 32B + GRITLM decomposition approach breaks a complex multi-hop query into sub-questions. While this can simplify the reasoning steps, it introduces a notable trade-off in retrieval specificity. As shown in Table~\ref{tab:your_label}, the decomposition-based approach demonstrates a larger gap between Hits@1 and Hits@5 compared to other methods. Specifically, the average gap from Hits@1 to Hits@5 for the decomposition approach is 13.95, which is significantly higher than GRITHopper’s 7.44, BeamRetriever's 6.57, and GRITLM's 8.45.

This substantial gap suggests that generated sub-queries often underspecify the necessary context, causing initial retrieval inaccuracies. While relevant passages appear among the top-$k$ retrieved documents, the first-ranked results are more likely to be off-target. We discuss this phenomenon with an example in more detail in Appendix \ref{app:specifity}.

By contrast, GRITHopper’s end-to-end differentiability preserves the full complexity of the query, yielding more specific embeddings that ensure relevant passages appear at the top, reducing the need for multiple autoregressive steps.

\section{Conclusion}
We introduced \textbf{GRITHopper-7B}, a novel multi-hop dense retrieval model that achieves state-of-the-art performance across both in-domain and out-of-distribution datasets. By training on extensive multi-hop datasets in question-answering and fact-checking, GRITHopper-7B outperforms previous decomposition-based methods while maintaining the efficiency of dense encoders. Our study demonstrated that decomposition-free approaches like GRITHopper surpass decomposition-based methods in multi-hop retrieval tasks due to better query specificity and reduced computational overhead. GRITHopper generalizes exceptionally well on out-of-distribution benchmarks, confirming its robustness across diverse datasets. We found that integrating causal language modeling with embedding training substantially enhances dense retrieval performance compared to embedding-only training. Additionally, incorporating post-retrieval language modeling by including final answers further improves the model’s ability to retrieve relevant passages, while causal negatives lead to stronger distractor but worse open retrieval performance. We have demonstrated how its generative training enables GRITHopper for end-to-end retrieval, outperforming previous state-of-the-art methods. We release GRITHopper-7B to the community as a resource for future research in natural language processing tasks requiring complex reasoning and retrieval capabilities.

%--> shown that casual language modeling ability for long reasoning helps improve multi-hop dense retrieval performance. Adding outside information, which we coin post retrieval 

\section{Limitations}
Despite its state-of-the-art performance, \textbf{GRITHopper-7B} has several limitations:

\begin{itemize}
    \item \textbf{Scalability Challenges for Large Corpora:} 
    While GRITHopper efficiently handles open-domain multi-hop retrieval, the reliance on pre-computed dense embeddings limits its scalability for extremely large corpora. The computational cost of creating and maintaining dense representations for frequent updates remains for a 7B model significant.
    
    \item \textbf{Dependency on High-Quality Hard Negatives:} 
    GRITHopper relies on effective hard negative mining to train contrastive objectives. This dependency may limit its applicability in domains or datasets lacking high-quality distractor annotations or the ability to mine suitable negatives. This is something we especially observe in reward learning, where there are substantial performance drops on datasets where we lack information on answers and sub-questions (like Fact-Checking) to determine which makes a passage irrelevant or relevant.  

    \item \textbf{Computational Overhead for Training:} 
    The integration of both embedding and generative objectives requires substantial GPU resources (e.g. 8 $\times$ A100-80GB GPUs). This makes GRITHopper less accessible for research groups with limited computational resources.

    \item \textbf{Sensitivity to Dataset Characteristics:} 
    GRITHopper performs exceptionally well on multi-hop tasks with well-defined retrieval chains (e.g., MuSiQue, HoVer). However, its performance on tasks with noisier or less structured retrieval chains (e.g., conversational QA) remains untested, highlighting potential brittleness to dataset variability.

    \item \textbf{Multi-Hop Dense Retrieval Model}
    Since, in contrast to GRITLM, we do not train on (retrieval-independent) instruction datasets in parallel, we do not expect that the model will perform well on generation on other tasks. Thus, our model is intended only for decomposition-free multi-hop dense retrieval.

%\item \textbf{Limited Exploration of Larger Model Architectures in Decomposition-free Multi-Hop Retrieval:}
%Prior decomposition-free multi-hop retrieval methods primarily relied on BERT-based architectures, constraining their capability to leverage larger, decoder-based language models. GRITLM, while utilizing a decoder-based model, was only trained for single-step retrieval in 2-hop scenarios, thereby limiting its effectiveness for complex multi-hop retrieval tasks requiring multiple context-driven steps . GRITHopper addresses this gap as the first decoder-based multi-hop dense retrieval model explicitly trained for deeper multi-hop problems. This is why GRITLM is quite good at the first hop (see Figure 1) but can not handle context at later hops. However, attempts to scale previous decomposition-free multi-hop retrieval BERT-based models resulted in overfitting and diminished performance (see Appendix \ref{app:beam}). Consequently, our work focuses extensively on ablation studies exploring incremental additions of auxiliary signals during training, transitioning from simple contrastive fine-tuning of GRITLM to incorporating causal language modeling objectives. A key limitation is the scope of experimentation; while our method represents an initial step towards utilizing decoder-based models for multi-hop retrieval, alternative methods remain unexplored and warrant further investigation.

\item \textbf{Absence of Directly Comparable Baselines for Decoder-based Multi-Hop Dense Retrieval:}

A central challenge in evaluating our approach arises from the lack of directly comparable baselines. Previous models either (a) employed decoder-based architectures but focused solely on single-hop retrieval for maximum 2-hop problems using only the question and no further context (e.g., GRITLM, \cite[p.~48]{muennighoff2024generative}), or (b) addressed multi-hop retrieval problems but exclusively utilized BERT-based architectures (e.g., MDR, BeamRetriever). As GRITHopper represents the first decoder-based model explicitly designed for decomposition-free multi-hop dense retrieval tasks, direct comparisons to prior work are inherently constrained. To address this, we (a) fine-tuned GRITLM on multi-hop datasets to establish a relevant decoder-based baseline and (b) from there conducted comprehensive ablation studies to clearly quantify and isolate the effects of each component within GRITHopper’s design. Furthermore, (c) we increased the size of encoder models (e.g. DeBERTaXL) for previous decomposition-free multi-hop retrieval models, which resulted in overfitting and diminished performance (see Appendix \ref{app:beam}).

\item \textbf{Limited Exploration of End-to-End Retrieval Dynamics:}
While GRITHopper enables end-to-end retrieval with generative control (e.g., stopping and reranking), these generative behaviors are not jointly optimized with dense retrieval. In particular, the highest stopping accuracy (75\%) is achieved at later checkpoints, whereas the model selected for reporting is chosen based on peak dense retrieval performance, resulting in a slightly lower stopping accuracy of 71.22\%. This reflects an inherent misalignment between optimal generative and embedding objectives. To contextualize this trade-off, we provide a detailed analysis of stopping behavior (early stopping vs. overshooting) and per-hop latency in the appendix. Future work should explore whether scaling the training data or explicitly aligning generative and retrieval objectives can further close this gap.
\end{itemize}
\section{Ethics}
The development and deployment of \textbf{GRITHopper-7B} raise two key ethical considerations. First, the model's reliance on large-scale datasets introduces the risk of propagating biases present in the training data  \citep{prakash-lee-2023-layered, Schramowski2022}, potentially leading to skewed retrieval outcomes or amplification of misinformation. Additionally, the open-domain nature of the retrieval task heightens the risk of retrieving sensitive or harmful content, which could pose challenges in privacy and content moderation. Second, GRITHopper’s decomposition-free approach reduces interpretability compared to methods that produce intermediate outputs, making it harder to explain and trust its decisions in high-stakes scenarios.

\section{Acknowledgement}
This work has been funded by the LOEWE initiative (Hesse, Germany) within the emergenCITY center (Grant Number: LOEWE/1/12/519/03/05.001(0016)/72). Furthermore, this research work has been funded by the German Federal Ministry of Education and Research and the Hessian Ministry of Higher Education, Research, Science and the Arts within their joint support of the National Research Center for Applied
Cybersecurity ATHENE. The model training was supported by a compute grant at the 42 supercomputer provided by hessian.AI (HMD: S-DIW04/0013/003; BMBF: 01IS22091). Additionally, we gratefully acknowledge the support of Microsoft with a grant for access to OpenAI GPT models via the Azure cloud (Accelerate Foundation Model Academic Research).

\bibliography{anthology,custom}

@inproceedings{warner2024smarterbetterfasterlonger,
    title = "Smarter, Better, Faster, Longer: A Modern Bidirectional Encoder for Fast, Memory Efficient, and Long Context Finetuning and Inference",
    author = {Warner, Benjamin  and
      Chaffin, Antoine  and
      Clavi{\'e}, Benjamin  and
      Weller, Orion  and
      Hallstr{\"o}m, Oskar  and
      Taghadouini, Said  and
      Gallagher, Alexis  and
      Biswas, Raja  and
      Ladhak, Faisal  and
      Aarsen, Tom  and
      Adams, Griffin Thomas  and
      Howard, Jeremy  and
      Poli, Iacopo},
    editor = "Che, Wanxiang  and
      Nabende, Joyce  and
      Shutova, Ekaterina  and
      Pilehvar, Mohammad Taher",
    booktitle = "Proceedings of the 63rd Annual Meeting of the Association for Computational Linguistics (Volume 1: Long Papers)",
    month = jul,
    year = "2025",
    address = "Vienna, Austria",
    publisher = "Association for Computational Linguistics",
    url = "https://aclanthology.org/2025.acl-long.127/",
    doi = "10.18653/v1/2025.acl-long.127",
    pages = "2526--2547",
    ISBN = "979-8-89176-251-0"
}

@inproceedings{
asai2024selfrag,
title={Self-{RAG}: Learning to Retrieve, Generate, and Critique through Self-Reflection},
author={Akari Asai and Zeqiu Wu and Yizhong Wang and Avirup Sil and Hannaneh Hajishirzi},
booktitle={The Twelfth International Conference on Learning Representations},
year={2024},
url={https://openreview.net/forum?id=hSyW5go0v8}
}

@inproceedings{
guan2024amor,
title={{AMOR}: A Recipe for Building Adaptable Modular Knowledge Agents Through Process Feedback},
author={Jian Guan and Wei Wu and zujie wen and Peng Xu and Hongning Wang and Minlie Huang},
booktitle={The Thirty-eighth Annual Conference on Neural Information Processing Systems},
year={2024},
url={https://openreview.net/forum?id=jImXgQEmX3}
}

@inproceedings{
muennighoff2024generative,
title={Generative Representational Instruction Tuning},
author={Niklas Muennighoff and Hongjin SU and Liang Wang and Nan Yang and Furu Wei and Tao Yu and Amanpreet Singh and Douwe Kiela},
booktitle={The Thirteenth International Conference on Learning Representations},
year={2025},
url={https://openreview.net/forum?id=BC4lIvfSzv}
}

@article{radford2019language,
  title        = {Language Models are Unsupervised Multitask Learners},
  author       = {Radford, Alec and Wu, Jeffrey and Child, Rewon and Luan, David and Amodei, Dario and Sutskever, Ilya},
  journal      = {OpenAI Blog},
  volume       = {1},
  number       = {8},
  pages        = {9},
  year         = {2019},
  url          = {https://cdn.openai.com/better-language-models/language_models_are_unsupervised_multitask_learners.pdf}
}

@article{liu2019robertarobustlyoptimizedbert,
      title={RoBERTa: A Robustly Optimized BERT Pretraining Approach}, 
      author={Yinhan Liu and Myle Ott and Naman Goyal and Jingfei Du and Mandar Joshi and Danqi Chen and Omer Levy and Mike Lewis and Luke Zettlemoyer and Veselin Stoyanov},
      year={2019},
      eprint={1907.11692},
      archivePrefix={arXiv},
      primaryClass={cs.CL},
      journal={arXiv:1907.11692},      
      url={https://arxiv.org/abs/1907.11692}, 
}

@inproceedings{
he2021debertav3,
title={De{BERT}aV3: Improving De{BERT}a using {ELECTRA}-Style Pre-Training with Gradient-Disentangled Embedding Sharing},
author={Pengcheng He and Jianfeng Gao and Weizhu Chen},
booktitle={The Eleventh International Conference on Learning Representations},
address={Kigali, Rwanda},
year={2023},
url={https://openreview.net/forum?id=sE7-XhLxHA}
}

@inproceedings{
khot2023decomposed,
title={Decomposed Prompting: A Modular Approach for Solving Complex Tasks},
author={Tushar Khot and Harsh Trivedi and Matthew Finlayson and Yao Fu and Kyle Richardson and Peter Clark and Ashish Sabharwal},
booktitle={The Eleventh International Conference on Learning Representations },
year={2023},
url={https://openreview.net/forum?id=_nGgzQjzaRy}
}

@inproceedings{
zelikman2022star,
title={{ST}aR: Bootstrapping Reasoning With Reasoning},
author={Eric Zelikman and Yuhuai Wu and Jesse Mu and Noah Goodman},
booktitle={Advances in Neural Information Processing Systems},
editor={Alice H. Oh and Alekh Agarwal and Danielle Belgrave and Kyunghyun Cho},
year={2022},
url={https://openreview.net/forum?id=_3ELRdg2sgI}
}

@article{schnitzler2024morehopqamultihopreasoning,
      title={MoreHopQA: More Than Multi-hop Reasoning}, 
      author={Julian Schnitzler and Xanh Ho and Jiahao Huang and Florian Boudin and Saku Sugawara and Akiko Aizawa},
      year={2024},
      eprint={2406.13397},
      journal={arXiv:2406.13397},      
      archivePrefix={arXiv},
      primaryClass={cs.CL},
      url={https://arxiv.org/abs/2406.13397}, 
}

@inproceedings{
asai2020learning,
title={Learning to Retrieve Reasoning Paths over Wikipedia Graph for  Question Answering},
author={Akari Asai and Kazuma Hashimoto and Hannaneh Hajishirzi and Richard Socher and Caiming Xiong},
booktitle={International Conference on Learning Representations},
url={https://openreview.net/forum?id=SJgVHkrYDH},
year={2020}
}

@inproceedings{shao-etal-2023-enhancing,
    title = "Enhancing Retrieval-Augmented Large Language Models with Iterative Retrieval-Generation Synergy",
    author = "Shao, Zhihong  and
      Gong, Yeyun  and
      Shen, Yelong  and
      Huang, Minlie  and
      Duan, Nan  and
      Chen, Weizhu",
    editor = "Bouamor, Houda  and
      Pino, Juan  and
      Bali, Kalika",
    booktitle = "Findings of the Association for Computational Linguistics: EMNLP 2023",
    month = dec,
    year = "2023",
    address = "Singapore",
    publisher = "Association for Computational Linguistics",
    url = "https://aclanthology.org/2023.findings-emnlp.620",
    doi = "10.18653/v1/2023.findings-emnlp.620",
    pages = "9248--9274",
}

@inproceedings{gao2021scaling,
    title = "Scaling Deep Contrastive Learning Batch Size under Memory Limited Setup",
    author = "Gao, Luyu  and
      Zhang, Yunyi  and
      Han, Jiawei  and
      Callan, Jamie",
    editor = "Rogers, Anna  and
      Calixto, Iacer  and
      Vuli{\'c}, Ivan  and
      Saphra, Naomi  and
      Kassner, Nora  and
      Camburu, Oana-Maria  and
      Bansal, Trapit  and
      Shwartz, Vered",
    booktitle = "Proceedings of the 6th Workshop on Representation Learning for NLP (RepL4NLP-2021)",
    month = aug,
    year = "2021",
    address = "Online",
    publisher = "Association for Computational Linguistics",
    url = "https://aclanthology.org/2021.repl4nlp-1.31/",
    doi = "10.18653/v1/2021.repl4nlp-1.31",
    pages = "316--321"
}

@inproceedings{ma-etal-2024-ex,
    title = "{EX}-{FEVER}: A Dataset for Multi-hop Explainable Fact Verification",
    author = "Ma, Huanhuan  and
      Xu, Weizhi  and
      Wei, Yifan  and
      Chen, Liuji  and
      Wang, Liang  and
      Liu, Qiang  and
      Wu, Shu  and
      Wang, Liang",
    editor = "Ku, Lun-Wei  and
      Martins, Andre  and
      Srikumar, Vivek",
    booktitle = "Findings of the Association for Computational Linguistics: ACL 2024",
    month = aug,
    year = "2024",
    address = "Bangkok, Thailand",
    publisher = "Association for Computational Linguistics",
    url = "https://aclanthology.org/2024.findings-acl.556",
    doi = "10.18653/v1/2024.findings-acl.556",
    pages = "9340--9353",
}

@article{moreira2024nvretrieverimprovingtextembedding,
      title={NV-Retriever: Improving text embedding models with effective hard-negative mining}, 
      author={Gabriel de Souza P. Moreira and Radek Osmulski and Mengyao Xu and Ronay Ak and Benedikt Schifferer and Even Oldridge},
      year={2024},
      journal={arXiv:2407.15831},
      eprint={2407.15831},
      archivePrefix={arXiv},
      primaryClass={cs.IR},
      url={https://arxiv.org/abs/2407.15831}, 
}

@inproceedings{
tang2024multihoprag,
title={MultiHop-{RAG}: Benchmarking Retrieval-Augmented Generation for Multi-Hop Queries},
author={Yixuan Tang and Yi Yang},
booktitle={First Conference on Language Modeling},
year={2024},
url={https://openreview.net/forum?id=t4eB3zYWBK}
}

@inproceedings{muennighoff-etal-2023-mteb,
    title = "{MTEB}: Massive Text Embedding Benchmark",
    author = "Muennighoff, Niklas  and
      Tazi, Nouamane  and
      Magne, Loic  and
      Reimers, Nils",
    editor = "Vlachos, Andreas  and
      Augenstein, Isabelle",
    booktitle = "Proceedings of the 17th Conference of the European Chapter of the Association for Computational Linguistics",
    month = may,
    year = "2023",
    address = "Dubrovnik, Croatia",
    publisher = "Association for Computational Linguistics",
    url = "https://aclanthology.org/2023.eacl-main.148",
    doi = "10.18653/v1/2023.eacl-main.148",
    pages = "2014--2037",
}

@inproceedings{
    thakur2021beir,
    title={{BEIR}: A Heterogeneous Benchmark for Zero-shot Evaluation of Information Retrieval Models},
    author={Nandan Thakur and Nils Reimers and Andreas R{\"u}ckl{\'e} and Abhishek Srivastava and Iryna Gurevych},
    booktitle={Thirty-fifth Conference on Neural Information Processing Systems Datasets and Benchmarks Track (Round 2)},
    year={2021},
    url={https://openreview.net/forum?id=wCu6T5xFjeJ}
}

@inproceedings{
zhang2024generative,
title={Generative Verifiers: Reward Modeling as Next-Token Prediction},
author={Lunjun Zhang and Arian Hosseini and Hritik Bansal and Mehran Kazemi and Aviral Kumar and Rishabh Agarwal},
booktitle={The 4th Workshop on Mathematical Reasoning and AI at NeurIPS'24},
year={2024},
url={https://openreview.net/forum?id=CxHRoTLmPX}
}

@inproceedings{yao2023react,
  title = {{ReAct}: Synergizing Reasoning and Acting in Language Models},
  author = {Yao, Shunyu and Zhao, Jeffrey and Yu, Dian and Du, Nan and Shafran, Izhak and Narasimhan, Karthik and Cao, Yuan},
  booktitle = {International Conference on Learning Representations (ICLR) },
  year = {2023},
  html = {https://arxiv.org/abs/2210.03629},
}

@inproceedings{
lee2025nvembed,
title={{NV}-Embed: Improved Techniques for Training {LLM}s as Generalist Embedding Models},
author={Chankyu Lee and Rajarshi Roy and Mengyao Xu and Jonathan Raiman and Mohammad Shoeybi and Bryan Catanzaro and Wei Ping},
booktitle={The Thirteenth International Conference on Learning Representations},
year={2025},
address={Singapore},
url={https://openreview.net/forum?id=lgsyLSsDRe}
}

@inproceedings{zhang-etal-2024-end,
    title = "End-to-End Beam Retrieval for Multi-Hop Question Answering",
    author = "Zhang, Jiahao  and
      Zhang, Haiyang  and
      Zhang, Dongmei  and
      Yong, Liu  and
      Huang, Shen",
    editor = "Duh, Kevin  and
      Gomez, Helena  and
      Bethard, Steven",
    booktitle = "Proceedings of the 2024 Conference of the North American Chapter of the Association for Computational Linguistics: Human Language Technologies (Volume 1: Long Papers)",
    month = jun,
    year = "2024",
    address = "Mexico City, Mexico",
    publisher = "Association for Computational Linguistics",
    url = "https://aclanthology.org/2024.naacl-long.96",
    doi = "10.18653/v1/2024.naacl-long.96",
    pages = "1718--1731",
}

@inproceedings{
xiong2021answering,
title={Answering Complex Open-Domain Questions with Multi-Hop Dense Retrieval},
author={Wenhan Xiong and Xiang Li and Srini Iyer and Jingfei Du and Patrick Lewis and William Yang Wang and Yashar Mehdad and Scott Yih and Sebastian Riedel and Douwe Kiela and Barlas Oguz},
booktitle={International Conference on Learning Representations},
year={2021},
url={https://openreview.net/forum?id=EMHoBG0avc1}
}

@inproceedings{huang-chang-2023-towards,
    title = "Towards Reasoning in Large Language Models: A Survey",
    author = "Huang, Jie  and
      Chang, Kevin Chen-Chuan",
    editor = "Rogers, Anna  and
      Boyd-Graber, Jordan  and
      Okazaki, Naoaki",
    booktitle = "Findings of the Association for Computational Linguistics: ACL 2023",
    month = jul,
    year = "2023",
    address = "Toronto, Canada",
    publisher = "Association for Computational Linguistics",
    url = "https://aclanthology.org/2023.findings-acl.67",
    doi = "10.18653/v1/2023.findings-acl.67",
    pages = "1049--1065",
}

@inproceedings{prakash-lee-2023-layered,
    title = "Layered Bias: Interpreting Bias in Pretrained Large Language Models",
    author = "Prakash, Nirmalendu  and
      Lee, Roy Ka-Wei",
    editor = "Belinkov, Yonatan  and
      Hao, Sophie  and
      Jumelet, Jaap  and
      Kim, Najoung  and
      McCarthy, Arya  and
      Mohebbi, Hosein",
    booktitle = "Proceedings of the 6th BlackboxNLP Workshop: Analyzing and Interpreting Neural Networks for NLP",
    month = dec,
    year = "2023",
    address = "Singapore",
    publisher = "Association for Computational Linguistics",
    url = "https://aclanthology.org/2023.blackboxnlp-1.22",
    pages = "284--295",
}

@Article{Schramowski2022,
author={Schramowski, Patrick
and Turan, Cigdem
and Andersen, Nico
and Rothkopf, Constantin A.
and Kersting, Kristian},
title={Large pre-trained language models contain human-like biases of what is right and wrong to do},
journal={Nature Machine Intelligence},
year={2022},
month={Mar},
day={01},
volume={4},
number={3},
pages={258-268},
issn={2522-5839},
doi={10.1038/s42256-022-00458-8},
url={https://doi.org/10.1038/s42256-022-00458-8}
}
\bibliographystyle{acl_natbib}

\appendix

\section{Baselines} \label{app:base}
\subsection{Beam Retriever}
\label{app:beam}
The Beam Retriever \citep{zhang-etal-2024-end} employs a cross-encoder architecture and relies on beam search to determine the number of steps required for retrieving multi-hop evidence. Unlike methods that have a predetermined number of computations, the Beam Retriever dynamically expands or shrinks the retrieval process, which is why the authors train with a Batch Size of $1$. Because large-scale parallelization on GPUs requires a uniform number of computations, this variability makes batching and distributed training for the model infeasible. Attempting to scale the Beam Retriever beyond DeBERTa-Base results in both performance degradation in open-retrieval and over-fitting on the distractor setting while facing dramatically increased training times. 
We tested ModerdBert Large \cite{warner2024smarterbetterfasterlonger}, DeBerta Large, DeBerta XL and the DeBerta base variant of the original paper. As highlighted in Table~\ref{tab:beamretriever_results}, we find that larger models, while achieving substantial performance improvements in the distractor setting, drop in performance in open retrieval on the same dataset. Showcasing the overfitting tendency to only train on distractors.

\begin{table}[h]
    \centering
    \small
    \begin{tabular}{lcc}
        \toprule
        \textbf{Model} & \multicolumn{2}{c}{\textbf{MuSiQue}} \\
        & \textbf{Distractor} & \textbf{Open Retrieval} \\
        \midrule
        DeBERTa Base  & 81.78 & \textbf{62.80} \\
        DeBERTa Large & \textbf{85.10} & 61.90 \\
        DeBERTa XL    & 72.36 & 58.24 \\
        ModernBert Large & 74.53 & 60.06 \\
        \bottomrule
    \end{tabular}
    \caption{BeamRetriever Performance on MuSiQue Distractor vs. MuSiQue Open Retrieval}
    \label{tab:beamretriever_results}
\end{table}

\subsection{MDR}
Multi-Hop Dense Retrieval (MDR) \citep{xiong2021answering} is natively designed for exactly two-hop retrieval. Efforts to extend MDR to more than two hops by adapting the loss function, as suggested by \citet{ma-etal-2024-ex}, led to instabilities in our experiments, including scenarios where the model's embeddings collapse. Since MDR’s loss is computed at the sample level, adapting it for varying hop lengths becomes non-trivial. These complexities, combined with the need to maintain large batch sizes for good generalization, hindered scaling to larger models or additional hops.

We train MDR on 8 $\times$ A100-80GB GPUs and find that batch size must decrease as model size grows. For instance, we can use a batch size of $16 \times 8$ for base models, $8 \times 8$ for roberta/deberta large ones, and only $2 \times 8$ for the largest variant (DeBerta XL). This reduction in batch size likely impacts the model’s generalization capabilities. Table~\ref{tab:MDR} in the main paper shows that even scaling MDR to RoBERTa-Large yields only minor improvements, and attempts to go beyond this configuration or handle more than two hops fail due to the aforementioned instabilities. To remain fair to the original authors, we report MDR results that remain as close as possible to their original setup. Bringing MDR up to today’s standards would likely involve adopting modern embedding objectives with techniques like gradient caching and instruction-tuned LLM backbones approaches we have integrated in our ablations with GRITHopper, where combining generative and embedding training yields superior performance compared to contrastive-only baselines (like MDR).
\begin{table*}[h!]
\centering
\footnotesize % Further reduces font size for better spacing
\renewcommand{\arraystretch}{1.1} % Increases row height for readability
\setlength{\tabcolsep}{4pt} % Slightly increases column spacing
\definecolor{lightgray}{gray}{0.9} % Define a light gray color
\resizebox{\textwidth}{!}{%
\begin{tabular}{l|ccccc|ccccc|ccccc|}
\toprule
  \multirow{2}{*}{Model} & \multicolumn{5}{c|}{Hits@1} & \multicolumn{5}{c|}{Hits@5} & \multicolumn{5}{c|}{Hits@10} \\
  \cmidrule(lr){2-6} \cmidrule(lr){7-11} \cmidrule(lr){12-16}
   & 1 & 2 & 3 & 4 & \cellcolor{lightgray}\itshape Avg & 1 & 2 & 3 & 4 & \cellcolor{lightgray}\itshape Avg & 1 & 2 & 3 & 4 & \cellcolor{lightgray}\itshape Avg \\
\midrule

\multicolumn{16}{l}{\textbf{Comparison to other models on MuSiQue}} \\
GRITHopper (ours) & 93.09 & 74.93 & 55.19 & 32.10 & \cellcolor{lightgray}\itshape 75.48 & 99.75 & 95.86 & 86.44 & 58.02 & \cellcolor{lightgray}\itshape 93.22 & 99.88 & 97.77 & 93.05 & 71.36 & \cellcolor{lightgray}\itshape 96.03 \\

GRITLM  & 91.15 & 57.51 & 22.32 & 5.43 & \cellcolor{lightgray}\itshape 60.51 & 99.50 & 91.31 & 65.49 & 35.56 & \cellcolor{lightgray}\itshape 86.18 & 99.96 & 96.61 & 83.26 & 51.85 & \cellcolor{lightgray}\itshape 92.61 \\
Beam Retriever  & 88.75 & 60.70 & 30.73 & 12.84 & \cellcolor{lightgray}\itshape 62.80 & 95.45 & 85.40 & 65.84 & 41.48 & \cellcolor{lightgray}\itshape 82.85 & 97.02 & 90.44 & 77.25 & 51.60 & \cellcolor{lightgray}\itshape 88.07 \\
Qwen 32B + GRITLM decomposition & 82.62 & 45.72 & 13.91 & 1.48 & \cellcolor{lightgray}\itshape 51.06 & 95.45 & 76.25 & 36.05 & 13.09 & \cellcolor{lightgray}\itshape 72.19 & 96.69 & 82.91 & 46.61 & 17.78 & \cellcolor{lightgray}\itshape 77.39 \\
\midrule
\multicolumn{16}{l}{\textbf{MDR on MuSiQue}} \\
DeBerta Base & 62.43 & 20.60 & - & - & \cellcolor{lightgray}\itshape 41.52 & 79.98 & 40.67 & - & - & \cellcolor{lightgray}\itshape 60.32 & 85.52 & 49.28 & - & - & \cellcolor{lightgray}\itshape 67.40  \\
Deberta Large & 74.35 & 32.06 & - & - & \cellcolor{lightgray}\itshape 53.21 & 85.97 & 52.25 & - & - & \cellcolor{lightgray}\itshape 69.11 & 89.78 & 59.95 & - & - & \cellcolor{lightgray}\itshape 74.87  \\
XL DeBerta  & 87.05 & 48.37 & - & - & \cellcolor{lightgray}\itshape 67.71 & 96.07 & 75.42 & - & - & \cellcolor{lightgray}\itshape 85.75 & 97.60 & 82.75 & - & - & \cellcolor{lightgray}\itshape 90.17  \\
Roberta Large  & 86.06 & 50.19 & - & - & \cellcolor{lightgray}\itshape \textbf{68.12}& 95.32 & 76.71 & - & - & \cellcolor{lightgray}\itshape 86.02 & 96.40 & 82.42 & - & - & \cellcolor{lightgray}\itshape 89.41  \\

\midrule

\multicolumn{16}{l}{\textbf{MDR on all Datasets}} \\
Roberta Large  & 81.75 & 45.18 & - & - & \cellcolor{lightgray}\itshape 63.47 & 94.37 & 71.04 & - & - & \cellcolor{lightgray}\itshape 82.71 & 96.73 & 78.82 & - & - & \cellcolor{lightgray}\itshape 87.77 \\
\midrule

\end{tabular}%
}
\caption{MDR ablations on different backbone architecturs}
\label{tab:MDR}
\end{table*}

\subsection{Decompostion based approach}\label{sec:decomp}
\label{sec:appendix-decomposition}

As discussed in Section~\ref{sec:base}, our decomposition-based baseline uses a step-by-step query decomposition approach. Each complex multi-hop question is decomposed into simpler sub-questions, and at each step we retrieve supporting paragraphs and extract the relevant answer. 

We employ four prompt templates for decomposition:
\begin{enumerate}[noitemsep,topsep=0pt]
    \item \textbf{First-Hop Sub-Question Generation:} Generates the initial sub-question from the original multi-hop question.
    \item \textbf{Second-Hop (Next) Sub-Question Generation:} Generates the next sub-question given the original question and the previously answered sub-questions.
    \item \textbf{Third-Hop (Next) Sub-Question Generation:} Similar to second-hop but for the third hop.
    \item \textbf{Fourth-Hop (Next) Sub-Question Generation:} Similar to above, for the fourth hop.
\end{enumerate}

Finally, we have an \textbf{Answer Extraction Prompt}, used after retrieving paragraphs, to extract the answer snippet.

% Multi-figure style: side-by-side boxes
\begin{figure*}[t]
    \centering
    % Left Box: Decomposition Prompt
    \begin{minipage}[t]{0.48\textwidth} % Adjust width to fit two boxes
    \begin{tcolorbox}[title=Prompt B.1: Decomposition of next Sub-Question]
You are given a multi-hop question and the answers to previous sub-questions. Given this information, break down the multi-hop question into the next smaller sub-question that can be answered by retrieving information via a search engine.

    \texttt{(Few-shot Examples: Multi-hop question + previous answers)}

    \textbf{Input:}
    \begin{verbatim}
Multi-hop Question: {multi_hop_question}
Previous Sub-Questions and Answers: {history}
    \end{verbatim}

    \textbf{Output:}
    \begin{verbatim}
Next Sub-Question: {generated_sub_question}
    \end{verbatim}
    \end{tcolorbox}
    \end{minipage}
    \hfill
    % Right Box: Answer Extraction Prompt
    \begin{minipage}[t]{0.48\textwidth} % Adjust width to fit two boxes
    \begin{tcolorbox}[title=Prompt B.2: Answer Extraction]
    You are given a question and a paragraph that contains the answer. Extract the relevant part of the paragraph that answers the sub-question. Ensure that the answer is as concise and accurate as possible.

    \texttt{(Few-shot Examples: Question + Retrieved Paragraph)}

    \textbf{Input:}
    \begin{verbatim}
Question: {sub_question}
Retrieved Paragraph: {retrieved_paragraph}
    \end{verbatim}

    \textbf{Output:}
    \begin{verbatim}
Answer: {extracted_answer}
    \end{verbatim}
    \end{tcolorbox}
    \end{minipage}
    \caption{Decomposition and Answer Extraction Prompt Templates. Few-shot examples include similar multi-hop problems with previously answered sub-questions and answers, demonstrating a consistent step-by-step structure. We provide a custom decomposition instruction for the first hop and provide custom 4 few-shot samples for each additional hop.}
\end{figure*}

\paragraph{Note on Evaluation Fairness:} 
We evaluate retrieval performance at each hop by checking if the correct evidence appears within the top-$k$ retrieved paragraphs. This evaluation is independent of the sub-questions order. Thus, regardless of how a model decomposes the problem, the evaluation remains fair and consistent across all methods.

%\paragraph{Transparency of GPT4o experiments}
%We provide the code for our GPT4o experiments and \href{https://anonymous.4open.science/r/GritHopper-AF16/readme.md}{GPT4o generations as part of our anonymous GitHub Repository.} 

\paragraph{Evaluation.} 
For evaluation, we follow a standard hits@k metric at each hop. We compare all models on their ability to retrieve the correct evidence at hop $1$, then at hop $2$, and so forth. To ensure a fair comparison, we do not rely on the self-correctness of decomposition-based methods as they inherently involve autoregressive generation, which allows multiple retries. In contrast, our decomposition-free approach computes a single dense embedding per step, making it significantly more efficient. While self-correction could improve performance, it introduces additional inefficiencies, contradicting the goal of comparing methods under the most efficient setting. Importantly, decomposition-based methods already require separate models for generation and embedding, further increasing computational cost.

\section{Training of GRITHopper}
In this section, we describe how GRITHopper was trained and how we derived the used training setup.

\subsection{Hard Negative Mining and Curriculum Learning}\label{sec:neg}
\citet{moreira2024nvretrieverimprovingtextembedding} have shown that mining difficult hard negatives is essential for achieving good dense retrieval performance. We employ the strongest GRITHopper model from our preliminary experiments, which has only been trained with distractors as hard negatives, to search via dense search the most difficult examples across the entire dataset for our final training run. For datasets like MuSiQue that provide entire decompositions (sub-questions with sub-answers for each hop), we filter distractors that contain the sub-answer. For other datasets where we are not able to filter this way, we filter negatives that have a cosine similarity higher than 0.95 to the positive paragraph. We select 10 hard negatives for the contrastive loss for each positive sample and add the most difficult one to our generative loss. We find that this is essential for making the causal reward learning work. 
Initially, we employed a curriculum learning approach: after each epoch, we used the current model’s predictions to mine new negatives for the subsequent epoch. However, longer training (beyond two or more epochs) led to overfitting and hindered out-of-distribution performance. We also tried taking the model checkpoints from one epoch to mine negatives, and then re-initializing a fresh model with those mined negatives. This approach did prove beneficial and improved 4\% on MusiQue in preliminary experiments. 

\subsection{Causal vs Dense Retrieval Performance}
\label{app:causvsdense}
We find that when training on all datasets, the peek performance on causal performance is only reached after 3 times longer training than for optimal embedding performance, leading to overfitting. To not sacrifice embedding generalization, GritHopper on all datasets has, therefore, a slightly weaker end-to-end performance at $71.22$ than its MuSiQue Only version at $75$. We observe this also in the re-ranking performance which is significantly lower at $59.04$, and although extended training improves re-ranking to up to $76.78$, it still does not surpass the embedding performance while leading to overfitting on the dense retrieval task. 

\section{Detailed Dataset adaptations}\label{sec:detailedData}
We first discuss the evaluation dataset specifics for evaluation and then our Training Dataset construction.

\subsection{Detailed dataset statistics for Evaluation}
Detailed evaluation dataset statistics are shown in Table~\ref{tab:dataset_statisticsEVAL}. For open retrieval, we use \emph{all} paragraphs associated with each dataset as candidate negatives, without any pre-filtering d. This reflects a realistic open-domain multi-hop retrieval scenario, where a model must retrieve relevant evidence from a large corpus rather than from a small, curated distractor set. We deliberately do not add additional external negatives beyond the dataset-provided paragraphs. Doing so would substantially increase the candidate pool and render a comparison to cross-encoder-based methods such as BeamRetriever infeasible. In our setup, the resulting candidate pool ranges from approximately 2{,}000 passages for MoreHopQA to over 20{,}000 passages for Explainable FEVER. At this scale, cross-encoder-based retrieval becomes computationally prohibitive. Since BeamRetriever must score each (query, passage) pair independently at every hop, runtime grows linearly with the number of candidate passages. In practice, this leads to runtimes of up to \textbf{400 GPU hours} for a single dataset in the open retrieval setting, as discussed in Appendix~\ref{app:speed}. This limitation is inherent to cross-encoder architectures and not specific to our implementation. To mitigate this asymmetry and enable a fairer comparison of modeling capacity independent of scalability constraints, we additionally evaluate all methods in a \emph{distractor setting}, where each query is paired with a restricted set of candidate passages. These experiments allow cross-encoders to operate in their intended regime and serve as a controlled analysis of training behavior and discriminative ability. We report distractor results separately and use them primarily for ablations and diagnostic comparisons (e.g., analyzing reward modeling and stopping behavior), while treating open retrieval as the primary indicator of real-world applicability. Dense retrieval models such as GRITHopper and MDR encode the corpus once offline and rely on efficient nearest-neighbor search during inference, making open retrieval evaluation tractable at this scale. For this reason, open retrieval serves as our main evaluation setting, complemented by distractor-based experiments to ensure comparability and transparency across architectural classes.

\subsection{Training Dataset}
We use the entire dataset of MuSiQue, HotpotQA as well as Hover. In Hover and ExFever, however, we find that not all hops are multi-hop if we remove duplicated evidence in the same sample, resulting in some 1-hop problems. The 2WikiMultiHopQA consist of only 2 hop and 4 hop problems, as we have a large amount from 2 hop problems already from HotpotQA, we only take 4 hop problems from there to not further unbalance the length of hops. While the post-retrieval information for MultiHop Question answering is clear, for fact-checking, we adapt whether the claim is supportive or unsupportive as the final answer. From Hover, we only use supporting paragraphs as it has no refuted label, making incomplete/irrelevant as positives unsuitable for contrastive learning.
\begin{table}[ht]
\tiny
\centering
\renewcommand{\arraystretch}{1.2} % Adjust row spacing for readability
\setlength{\tabcolsep}{6pt} % Adjust column spacing for readability
\begin{tabular}{l|c|cccc}
\hline
\textbf{Dataset} & \textbf{Total Samples} & \multicolumn{4}{c}{\textbf{Samples Per Hop}} \\ 
\cline{3-6}
 & & \textbf{Hop 1} & \textbf{Hop 2} & \textbf{Hop 3} & \textbf{Hop 4} \\ \hline
MuSiQue & 19,938 & 0 & 14,376 & 4,387 & 1,175 \\
HoVer & 10,280 & 3,762 & 5,579 & 883 & 56 \\
HotpotQA & 90,447 & 0 & 90,447 & 0 & 0 \\
ExFever & 28,774 & 1,272 & 17,444 & 10,058 & 0 \\
2WikiMultiHopQA & 34,942 & 0 & 0 & 0 & 34,631 \\ \hline
\textbf{Total} & \textbf{184,070} & \textbf{5,034} & \textbf{127,846} & \textbf{15,328} & \textbf{35,862} \\ \hline
\end{tabular}
\caption{Training dataset statistics, including the total number of samples and the distribution of samples across different hop depths for each dataset. The final row shows the aggregate totals, providing an overview of the dataset scale when training across all datasets.}
\label{tab:dataset_construction}
\end{table}

\subsubsection{Open Evaluation statistics}
\begin{table}[ht]
\tiny
\centering
\renewcommand{\arraystretch}{0.4} % Adjust row spacing for readability
\setlength{\tabcolsep}{8pt} % Adjust column spacing for readability
\begin{tabular}{l|c|cccc}
\hline
\textbf{Dataset} & \textbf{total} & \multicolumn{4}{c}{\textbf{Samples Per Hop}} \\ 
\cline{3-6}
  & & \textbf{1} & \textbf{2} & \textbf{3} & \textbf{4} \\ \hline
MoreHopQA & 1,118 & 0 & 1,118 & 0 & 0 \\
ExFever & 8,038 &  166 & 4,671 & 3,201 & 0 \\
MuSiQue & 2,417 & 0 & 1,252 & 760 & 405 \\
MultiHopBench & 2,556 &  0 & 1,079 & 778 & 398 \\
Hover & 1,885 & 617 & 919 & 323 & 26 \\ \hline
\end{tabular}
\caption{Dataset statistics for the open retrieval evaluation setup. The table includes the number of multi-hop problems and the distribution of samples across different hop depths for each dataset.}
\label{tab:dataset_statisticsEVAL}
\end{table}

\newpage

In this section, we compare the computational complexity of a cross-encoder-based multi-hop retriever (e.g., Beam Retriever) and a dense bi-encoder-based multi-hop retriever (e.g., GRITHopper and MDR) under the scenario where both must consider the entire corpus of $P$ passages at each retrieval hop. This corresponds directly to the setting in our experiments, where the Beam Retriever processes all $P$ passages at every hop without a first-stage filter, resulting in prohibitively long runtimes.

\section{End-to-End Retrieval Analysis}
\label{app:diagnostics}
\begin{table*}[t]
    \centering
    \scriptsize
    \begin{tabular}{lcccc}
        \toprule
        \textbf{Model} &
        \textbf{Early Stops} &
        \textbf{Overshoots} &
        \textbf{Avg. Hops} &
        \textbf{Avg. Latency / Hop (s)} \\
        \midrule
        GRITHopper (end-to-end) & 347 & 25 & 2.41 & 0.29 \\
        BeamRetriever (end-to-end) & 945 & 0 & 1.83 & 0.41 \\
        \bottomrule
    \end{tabular}
    \caption{End-to-end retrieval diagnostics on MuSiQue. Early stops indicate premature termination before retrieving all required evidence. Overshoots indicate retrieval beyond the gold hop length. Latency is averaged per hop over all evaluated queries.}
    \label{tab:end2end_diagnostics}
\end{table*}

\begin{table}[t]
    \centering
    \small
    \begin{tabular}{lcccc}
        \toprule
        \textbf{Model} & \textbf{Hop 1} & \textbf{Hop 2} & \textbf{Hop 3} & \textbf{Hop 4} \\
        \midrule
        GRITHopper & 0.13 & 0.19 & 0.31 & 0.52 \\
        BeamRetriever & 0.25 & 0.27 & 0.46 & 0.66 \\
        \bottomrule
    \end{tabular}
    \caption{Average latency per hop (seconds). Dense retrieval scales more favorably with hop depth than cross-encoder retrieval.}
    \label{tab:latency_per_hop}
\end{table}

\section{Algorithm Dataset formatting}
\label{sec:algo}
\begin{algorithm}[H]
\caption{Dataset Construction for Multi-Hop Retrieval. For each multi-hop problem, the algorithm iterates through each hop (decomposition step). At each hop, it creates a contrastive sample consisting of the current retrieval prompt, a positive paragraph (supporting evidence), and a mined hard negative paragraph. Additionally, it generates a causal (generative) negative sample indicating the irrelevance of the mined negative paragraph. After processing all hops, the final generative positive sample includes the complete retrieval chain followed by the final answer. One random negative generative sample from the set of causal negatives is also selected to balance the dataset.}
\label{alg:dataset-construction}
\textbf{Input:} Multi-hop dataset $\mathcal{D} = \{(q, \mathcal{P}, a)\}$, where $q$ is the question, $\mathcal{P}$ is the set of paragraphs, $\mathcal{P}_s \subseteq \mathcal{P}$ are supporting paragraphs, and $a$ is the final answer.\\
\textbf{Output:} Generative samples $\mathcal{S}_g$, Contrastive samples $\mathcal{S}_r$.

\begin{algorithmic}[1]
\State Initialize $\mathcal{S}_g \gets \emptyset$, $\mathcal{S}_r \gets \emptyset$
\State Set instructions $Inst_Q$, $Inst_D$, and actions
\For{$(q, \mathcal{P}, a) \in \mathcal{D}$}
    \State $P \gets Inst_Q + q$ \\ \Comment{Initialize retrieval prompt}
    \State $\mathcal{S}_{neg} \gets \emptyset$
    \For{$i = 1$ to $|\mathcal{Q}_d|$} \\ \Comment{Iterate through decomposition steps $\mathcal{Q}_d$}
        \State $P \gets \mathcal{Q}_d[i]$
        \State $D_{neg} \gets mine\_negative(P,\mathcal{P})$
        \State  $D_{pos} \gets \mathcal{P}_s[i]$
        \State $\mathcal{S}_r \gets \mathcal{S}_r \cup (P, D_{pos}, D_{neg})$
        \State $P_{neg} \gets P + \texttt{Document: } D_{neg}$
        \State $P_{neg} \gets P_{neg} + \texttt{Eval(neg)}$
        \State $\mathcal{S}_{neg} \gets \mathcal{S}_{neg} \cup P_{neg} $
        \\ \Comment{next continue with positive chain}
        \State $P \gets P + \texttt{Document: } D_{pos}$
        \State $P \gets P + \texttt{Eval(pos)}$
        \If{$i \neq |\mathcal{Q}_d|$} \Comment{Final step}
            \State $P \gets P +\texttt{Retr}$
        \EndIf
    \EndFor
        \State $P_{final} \gets P + \texttt{Answer: } a$
        \State $\mathcal{S}_g \gets \mathcal{S}_g \cup P_{final}$
        \State $\mathcal{S}_g \gets \mathcal{S}_g \cup \texttt{random\_select}(\mathcal{S}_{neg})$
        \\ \Comment{to balance positive and negatives}
\EndFor
\State \Return $\mathcal{S}_g, \mathcal{S}_r$
\end{algorithmic}
\end{algorithm}
\section{Inference Speed Comparison}\label{app:speed}
\label{sec:complexity_appendix}
In this section, we compare the inference performance of GRITHopper to other methods. For decomposition-free approaches such as BeamRetriever, we analyze both computational complexity and actual inference time. In contrast, for decomposition-based methods (e.g., GPT-4.0 or Qwen-32B combined with a GRITLM retriever), which involve multiple model calls and variable generative steps, we report only the empirical inference time.

\subsection{Inference Time of Decomposition-based Methods}
\begin{table*}[h]
\centering
    \small
\begin{tabular}{lccc}
\toprule
\textbf{Method} & \textbf{Hardware} & \textbf{Query-Time Processing} & \textbf{Time (Full MuSiQue)} \\
\midrule
GRITHopper & 1× A100 80GB & Single pass & 6 min 42 s \\
GPT-4.0 (4-shot) & API + GRITLM (1× A100) & Retrieval + LLM API & 1 h 24 min 18 s \\
Qwen-32B (4-shot) & 2× A100 80GB & Retrieval + LLM (vLLM + KV cache) & 1 h 1 min 11 s \\
\bottomrule
\end{tabular}
\caption{Query-time inference comparison on the full MuSiQue dataset. Note that decomposition-based setups involve additional compute and API usage.}
\label{tab:decomp_inference_time}
\end{table*}
While our primary focus is on decomposition-free methods, we also provide inference time comparisons with decomposition-based approaches to contextualize the efficiency of GRITHopper.

Unlike decomposition-free methods, which encode a query and retrieve in a single pass, decomposition-based methods invoke the retriever and generator in multiple stages, typically per sub-question. As such, a theoretical complexity analysis is not meaningful due to the variability introduced by multiple model invocations and API-based generation. Instead, we report empirical inference times on the full MuSiQue dataset.

Table~\ref{tab:decomp_inference_time} summarizes the total inference time. GRITHopper completes retrieval in under 7 minutes, while GPT-4.0 and Qwen-32B, both operating in a 4-shot setting, require over an hour. These measurements include all query-time computation but exclude shared index encoding costs, which are identical across methods using GRITLM-based retrieval.

Note that these comparisons are not strictly fair: decomposition-based methods rely on external APIs or multiple GPUs, making the setup and latency less controllable. 

\subsection{Complexity Analysis Decomposition-free Methods}

\noindent\textbf{Notation:}
\begin{itemize}[noitemsep,nolistsep]
    \item $Q$: Number of queries.
    \item $H$: Average number of hops per query.
    \item $P$: Total number of passages in the corpus.
    \item $L_q$: Length (in tokens) of the query plus previously retrieved context.
    \item $L_p$: Length (in tokens) of a passage.
    \item $\mathcal{C}_{model}(L)$: Compute cost of a single forward pass on an input of length $L$.
    \item $\mathcal{C}_{search}(P)$: Compute cost of searching $P$ pre-encoded embeddings (sub-linear in $P$ using ANN indexes).
\end{itemize}

\subsubsection{Cross-Encoder (Beam Retriever)}
The cross-encoder must re-encode each passage together with the query at every hop. Without any pre-retrieval pruning, it compares against all $P$ passages each time:
\[
O(Q \cdot H \cdot P \cdot \mathcal{C}_{model}(L_q + L_p)).
\]
Since every passage is processed through the cross-encoder at every hop, runtime grows linearly with $P$ and $H$. For large $P$, this becomes extremely time-consuming (e.g., hundreds of hours).

\subsubsection{Dense Bi-Encoder (GRITHopper)}
Dense retrieval encodes all $P$ passages \emph{once} offline:
\[
O(P \cdot \mathcal{C}_{model}(L_p)).
\]
At inference time, each hop only requires encoding the query and performing a vector search over $P$:
\[
O(Q \cdot H \cdot [\mathcal{C}_{model}(L_q) + \mathcal{C}_{search}(P)]).
\]
Because the passages are already encoded, the cost per hop is dominated by a single query encoding and efficient similarity search. This typically takes orders of magnitude less time than re-encoding $P$ passages at every hop.

\subsubsection{Discussion}
Under identical conditions, considering all $P$ passages at each hop, the Beam Retriever’s complexity grows as $O(Q \cdot H \cdot P)$ with a high per-pass token cost, resulting in very long runtimes (e.g., over 400 hours in our ExFever open-retrieval experiments). In contrast, GRITHopper amortizes passage encoding and relies on fast search structures, completing the same task in 8 minutes and 20 seconds. This substantial practical difference in runtime reflects the asymptotic advantage of dense retrieval for large-scale, multi-hop scenarios.

\section{Loss of Specificity in Decomposition-based Methods} \label{app:specifity}
\begin{figure*}
    \centering
    \includegraphics[width=1.0\linewidth]{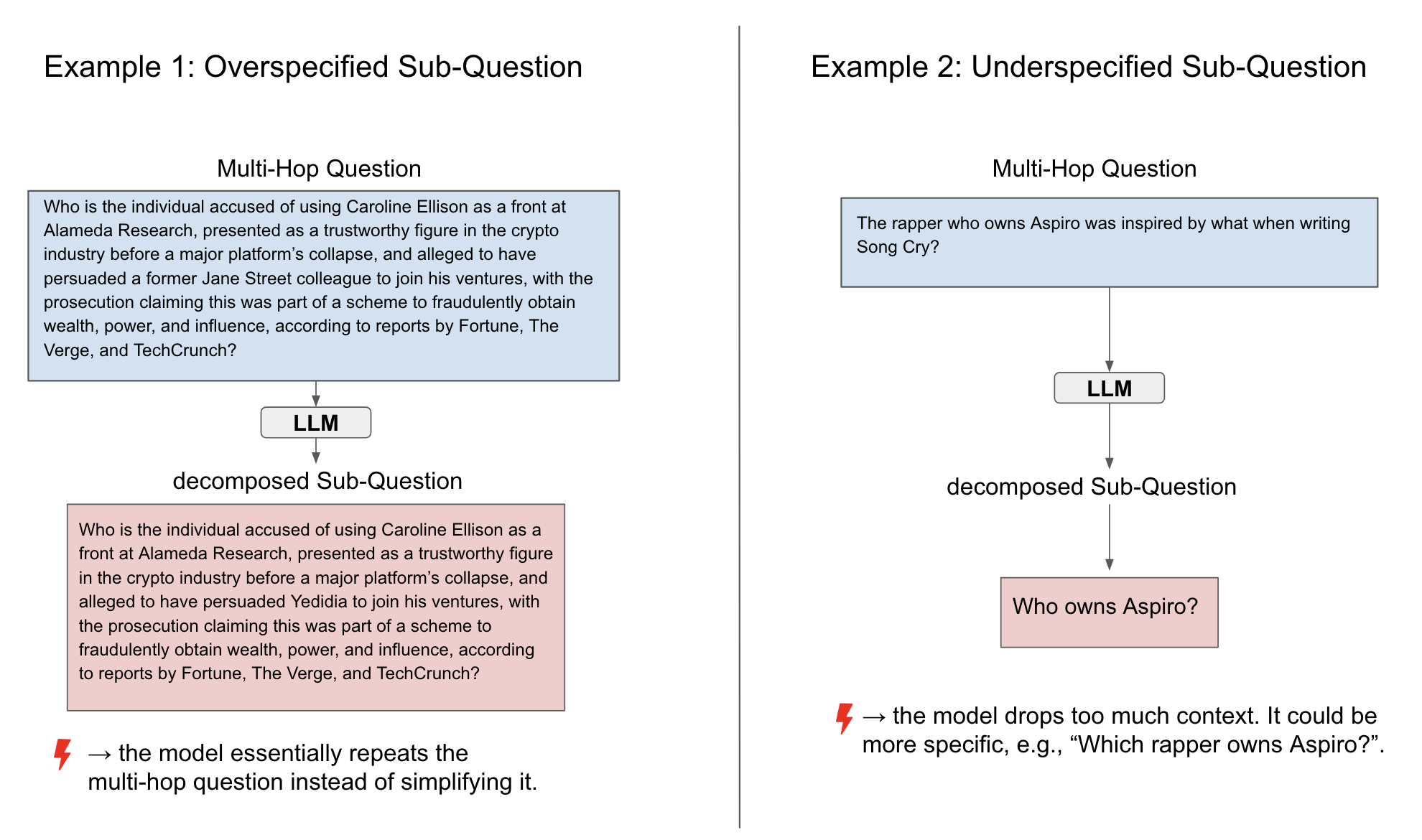}
    \caption{The decomposed queries from GPT4 and Qwen have a lower Hit@1 rate but get closer to our model at Hits@5, showing that the sub-questions are not necessarily wrong but are not specific enough to retrieve them at Hits@1. There are two reasons: (1) A subquestion is asking for too much information (left side), in other words, the sub-question is still multi-hop or (2) underspecified (right side), there is missing information (eg. entities) in the sub-question. Both lead to a lower hit rate at Hits@1 but a more competitive performance at higher Hits@k.}
    \label{fig:last}
\end{figure*}

See Figure \ref{fig:last}.

\section{Training Time Comparison}
See Table \ref{tab:trainingtime}.
\label{sec:trainingtime}
\begin{table*}[t]
    \centering
    \small

    \label{tab:training_time_comparison}
    \begin{tabular}{l c c c c c}
        \toprule
        \textbf{Model} & \textbf{Trained Epochs} & \textbf{Best Perf. Epoch} & \textbf{\# GPUs} & \textbf{Training Time (h)} & \textbf{Total GPU Hours} \\
        \midrule
        GritHopper (7B) & 5 & 1-2 & 8 & 181 & 1448 \\
        \midrule
        BeamRetriever DeBERTa XL & 10 (default: 20) & -$^{*}$ & 1 & 452 & 452 \\
        BeamRetriever DeBERTa Large & 20 & 14 & 1 & 289 & 289 \\
        BeamRetriever DeBERTa Base & 20 & 7 & 1 & 112 & 112 \\
        \bottomrule
    \end{tabular}%
        \caption{Training time comparison of different retrieval models trained on all datasets. The table shows the base model, the number of trained epochs, the best performance epoch, the number of GPUs used, the total training time in hours, and the total GPU hours (number of GPUs $\times$ training time). -$^{*}$ performance plateau was not reached.}
        \label{tab:trainingtime}

\end{table*}

\section{Training Setup}\label{app:trainingSetup}
GRITHopper-7B is trained on 8 $\times$ A100-80GB GPUs with a contrastive batch size of $2048$ using GradCache \cite{gao2021scaling} and a 256 batch size for the generative loss, like GRITLM in a Fully Sharded Data Parallel (FSDP) setup. We train all models for 5 epochs and select the best checkpoint via dense retrieval performance in the distractor setting.

\end{document}